\documentclass[useAMS,usenatbib]{mn2e}

\usepackage{epsfig}
\def\Lsun{L$_{\odot}$}
\newcommand{\Msun}{\mbox{$\rm M_{\odot}$}}

\title[Nuclear stellar discs in simulations]{Searching for nuclear
 stellar discs in simulations of star cluster mergers}

\author[E. Portaluri et al.]{E. Portaluri,$^{1}$\thanks{E-mail:
elisa.portaluri@studenti.unipd.it} E. M. Corsini,$^{1,2,}$
L. Morelli,$^{1,2}$ M. Hartmann,$^{3}$ E. Dalla Bont\`a,$^{1,2}$ \and
Victor P. Debattista,$^{4}$ and A. Pizzella$^{1,2}$\\
\\$^{1}$Dipartimento di Fisica e Astronomia `G. Galilei', 
 Universit\`a di Padova, vicolo dell'Osservatorio 3, I-35122 Padova, Italy
\\$^{2}$INAF--Osservatorio Astronomico di Padova, vicolo
 dell'Osservatorio 5, I-35122 Padova, Italy
\\$^{3}$Astronomisches Rechen-Institut, Zentrum f\"ur Astronomie 
 der Universit\"at Heidelberg (ZAH), M\"onchhofstra{\ss}e 12-14,\\ 
 D-69120 Heidelberg, Germany
\\$^{4}$Jeremiah Horrocks Institute, University of Central 
 Lancashire, Preston PR1 2HE, UK}

\begin{document}

\date{Accepted 2013 April 26. Received 2013 April 26; 
in original form 2013 February 04}

\pagerange{\pageref{firstpage}--\pageref{lastpage}} \pubyear{2012}

\maketitle

\label{firstpage}

\begin{abstract}
 The nuclei of galaxies often host small stellar discs with
 scale-lengths of a few tens of parsecs and luminosities up to $10^7$
 \Lsun .
 To investigate the formation and  properties of nuclear stellar discs
 (NSDs), we look for their presence in a set of $N-$body simulations
 studying the dissipationless merging of multiple star clusters in
 galactic nuclei.  A few tens of star clusters with sizes and masses
 comparable to those of globular clusters observed in the Milky Way
 are accreted onto a pre-existing nuclear stellar component: either a
 massive super star cluster or a rapidly rotating, compact disc with
 a scale-length of a few parsecs, mimicing the variety of observed
 nuclear structures.
 Images and kinematic maps of the simulation time-steps are then built
 and analysed as if they were real and at the distance of the Virgo
 cluster.  We use the Scorza-Bender method to search for the presence
 of disc structures via photometric decomposition. In one case the
 merger remnant has all the observed photometric and kinematic
 properties of NSDs observed in real galaxies.
 This shows that current observations are consistent with most of the
 NSD mass being assembled from the migration and accretion of star
 clusters into the galactic centre.
 In the other simulation instead, we detect an elongated structure
 from the unsharp masked image, that does not develop the photometric
 or kinematic signature of a NSD. Thus, in the context of searches for
 a disc structure, the Scorza-Bender method is a robust and necessary
 tool.
\end{abstract}

\begin{keywords}
galaxies: evolution -- galaxies: formation -- 
galaxies: kinematics and dynamics -- galaxies: nuclei -- 
galaxies: photometry -- galaxies: structure
\end{keywords}

\section{Introduction}
\label{sec:introduction}

The nuclei of galaxies occupy a privileged position at the bottom of
the galactic potential well. Thus the morphology, dynamics, star
formation, and chemical enrichment provide important insight into the
evolutionary history of galaxies.  Indeed galactic nuclei record the
history of the material, whether coming from neighbouring galactic
regions or accreted through mergers, that has sunk to the dynamical
centre over the lifetime of the parent system.

Due to its superb spatial resolution, the {\em Hubble Space Telescope
  (HST)\/} has been used extensively to image the nuclei of nearby
galaxies.  Numerous surveys have been carried out to investigate their
structure on scales of 10-100 pc over a wide range of morphological
types. {\em HST} imaging has shown that about $20$ per cent of
elliptical \citep{Ledo2010} and spiral galaxies \citep{Pizzella2002}
host a nuclear stellar disc (NSD). NSDs are smaller (scale-lengths $h
\sim 10-50$ pc) and brighter ($\mu_{0,V}^0 \sim 16-19$ mag
arcsec$^{-2}$) than embedded discs in early-type galaxies
\citep{Scorza1995}, but they never dominate the light distribution of
the galactic nuclei, and locally contribute at most half the galaxy
surface brightness.  Interestingly, independent of the Hubble type of
their host galaxy, NSDs seem to follow the same relation between the
central surface-brightness and scale-length as the main discs of
lenticular and spiral galaxies and as the embedded discs of early-type
galaxies \citep{vandenBosch1998}.

Various methods have been used to detect NSDs.  The simplest, and
therefore traditional, method is the unsharp-mask technique to recover
non-circular structures in an image.  Alternatively, the Scorza-Bender
method \citep{Scorza1995} recovers the basic structural properties of
a NSD (central surface brightness, $I_0$, and the scale-length radius,
$h$, inclination, $i$, and major-axis position angle, PA) by
subtracting from the image the best exponential disc model, which is
generally assumed to be perfectly thin.

NSDs have received attention primarily as probes of the mass
distribution since they permit measurement of the mass of central
supermassive black holes (SBHs) through observation of their stellar
kinematics \citep[e.g.,][]{vandenBosch1996, Magorrian1999,
 Cretton1999}. However, the study of NSDs can also advance our
understanding of galaxy formation and evolution since their formation
and destruction reflects the assembly history of their host galaxies.
For example, being dynamically cool systems, NSDs are fragile and do
not survive a major merger.  Especially in the presence of SBHs of
comparable mass, they would be significantly affected by the
interaction with a second SBH carried by the merging galaxy
\citep{Ledo2010}. Therefore, the age of a NSD constrains the epoch of
the last major merger of the host galaxy.

Different mechanisms have been proposed to explain how NSDs form.
The capture of external gas accounts for the origin of the
counter-rotating NSD in NGC~4458 \citep{Morelli2004, Morelli2010}.
Indeed, the presence of a kinematically decoupled component is usually
interpreted as the signature of gas accretion or merging of satellite
galaxies \citep{Bertola1999a}.  An external origin is also invoked for
the gas which formed the stars of the nuclear disc in the early-type
spiral NGC~4698, which is rotating perpendicularly with respect to the
main galactic disc \citep{Bertola1999b, Pizzella2002, Corsini1999,
 Corsini2012}.
The secular infall of gas, perhaps via a barred potential, is an
alternative way to funnel pre-enriched material into the galactic
centre where it accumulates, dissipates, and forms stars.  This
scenario provides a natural explanation for the presence of NSDs in
barred galaxies (e.g., NGC~7332, \citealt{Seifert1996,
 Falcon-Barroso2004}).  In some cases a bar may even be destroyed in
the process, as has been proposed for NGC~4570, \citep{Scorza1998,
 vandenBosch_Emsellem1998, vandenBosch_etal1998, Krajnovic2004,
 Morelli2010} and NGC~4621 \citep{Silchenko1997}).

Examples of on-going formation via dissipation include the NSDs in
NGC~5845 \citep{Kormendy1994} and in NGC~4486A \citep{Kormendy2005}.
However, other NSDs are as old as the main body of the host galaxy
(NGC~4128, \citealt{Krajnovic2004}; NGC~4342,
\citealt{vandenBosch_etal1998}; NGC~4458, \citealt{Morelli2004};
NGC~4621, \citealt{Krajnovic2004}; NGC~4698, \citealt{Corsini2012}).
The stellar populations in NSDs have been studied in detail in only a
few more objects.  In NGC~4478 the NSD is younger, more metal-rich
and less over-abundant than the rest of the galaxy
\citep{Morelli2004}.  In contrast the NSD of NGC~5308 is made of a
younger and more metal-poor stellar population than the host galaxy
\citep{Krajnovic2004}. Both the NSD and bulge of NGC~4570 show an
intermediate-age stellar population although the NSD is more metal
rich \citep{Krajnovic2004}.

In some cases a NSD co-exists with a massive and dense nuclear star
cluster with typical $L_V \sim 10^6-10^7$ \Lsun\ and
effective radius of a few parsecs. These objects are observed in the
centre of both early- (e.g. NGC~4458, NGC~4478, and NGC~4570,
\citealt{Ferrarese2006, Ledo2010}) and late-type galaxies (e.g.
NGC~4206, \citealt{Seth2006}).

Despite recent progress, no satisfying explanation has been provided
for the deposition of the high gas densities that are needed to enable
an {\em in situ} formation of nuclear star clusters
\citep{Shlosman1989, Milosavljevic2004, Bekki2007, Emsellem2008}. An
alternative scenario for the build-up of nuclear star clusters is the
dissipationless coalescence of star clusters sinking to the bottom of
the potential well by dynamical friction \citep{Tremaine1975,
  Capuzzo1993, Lotz2001, Capuzzo2008a, Capuzzo2008b, Antonini2012,
  Antonini2013}.
\citet{Agarwal2011} showed that the rate at which the nuclear star
cluster grows by accreting young star clusters depends on their
formation rate, migration time, and dissolution time.  The resulting
compact nuclear star cluster is embedded in a more extended and
diffuse component resembling a small pseudo-bulge.  The size of such a
disc-like structure in models including the prompt dissolution of star
clusters matches that of the observed NSDs. This opens the possibility
that NSDs may be forming via star cluster accretion.
In addition, the observed scaling relations between the nuclear star
cluster masses and the velocity dispersion of their host spheroids and
between the sizes of nuclear star clusters and their luminosities can
be explained by a dissipationless scenario \citep{Antonini2013}.

Some nuclear star clusters have a complex structure possessing both a
spheroidal component and an elongated ring or disc component with a
scale-length of a few parsecs \citep{Seth2006, Seth2008}.
\citet[][hereafter H$11$]{Hartmann2011} and \citet{DeLorenzi2013}
found that no more than half the mass of the multi-component nuclear
star cluster in NGC~4244 could have been assembled in star clusters
that migrated and merged at the centre of the galaxy. They concluded
that the rest of the nuclear star cluster mass must have been produced
by {\em in situ} star formation.  In this paper we analyse some of the
$N-$body simulations of H$11$ to look for the presence of a NSD and to
test whether the dissipationless merger of star clusters is a viable
scenario for the formation and growth of NSDs (as opposed to nuclear
star clusters).  The details of the simulations are given in
Section~\ref{sec:simulations}.  Section~\ref{sec:analysis} describes
the analysis of their photometric and kinematic properties based on
mock images and kinematic maps of the simulated galaxies
(Section~\ref{sec:data}), including the search for NSDs
(Section~\ref{sec:unsharp}), the measurement of their structural
properties using the Scorza-Bender method
(Section~\ref{sec:decomposition}), and the computation of the rotation
parameter (Section~\ref{sec:rotation}).  Results and conclusions are
discussed in Section~\ref{sec:discussion} and \ref{sec:conclusions},
respectively.

\section{Simulations}
\label{sec:simulations}

The simulations used here have been presented already in H$11$.  Here
therefore we provide only a brief description of the simulations and
refer the reader to H$11$ for more details.  The two simulations we
use are referred to as A1 and A2 in H$11$ and we continue to refer to
them as such here.

We are interested in the evolution in the inner $\sim 100$ pc of
galaxies; we therefore neglect the dark matter halo.  Both simulations
live at the centre of a \citet{Hernquist1990} bulge:
\begin{equation}
\rho(r) = \frac{a M_{\rm b}}{2\pi r \left(r+a\right)^3},
\label{eqn:bulge}
\end{equation}
where $M_{\rm b}$ is the bulge mass and $a$ is the scale radius.  We
use $M_{\rm b} = 5\,\times\,10^9$ \Msun\ and $a = 1.7$ kpc.  The bulge
is populated by $3.5 \times 10^6$ particles with masses ranging from
$40$ \Msun\ at the centre to $3.9\,\times\,10^5$ \Msun\ further out
\citep{Sellwood2008}; their softening is related to their mass via
$\epsilon_p \propto m_p^{1/3}$, as shown in Fig.~8 of H$11$.

We set up model star clusters, ranging in mass from
$2\,\times\,10^5$ \Msun\ to $2\,\times\,10^6$ \Msun, using the
isotropic distribution function of a lowered polytrope with index
$n=2$:
\begin{equation}
f(x,v) \propto [-2 E(x,v)]^{1/2} - [-2 E_{\rm max}]^{1/2}
\label{eqn:polytropedf}
\end{equation}
where $(x,v)$ are the phase space coordinates, and $E$ is the energy.
This distribution function reproduces well the core profile of
observed star clusters, as shown in Fig.~9 of H$11$.  We produce
equilibrium models via the iterative procedure described in
\citet{Debattista2000}. We set up three such models, C3-C5, using the
naming scheme of H$11$.  Star cluster models have all particles ($4
\times 10^5$ for C3 model and $4 \times 10^4$ for C4-C5) of equal mass
($5.0$ \Msun\ for C3-C4 and $15$ \Msun\ for C5) and equal softening
($\epsilon =0.13$ pc). No black holes have been included in the star
cluster models. Table~\ref{tab:scs} lists the properties of the star
cluster models. The concentration $c$ is defined as $c \equiv
\log{(R_{\rm e}/R_{\rm c})}$ where $R_{\rm e}$ is the half mass radius
and $R_{\rm c}$ is the core radius, where the surface density drops to
half of the central.  The masses and sizes of the star clusters are
comparable to young massive star clusters in the Milky Way
\citep{Figer1999, Figer2002}.

\begin{table}
\caption{\small{Star clusters used in the simulations.}}
\label{tab:scs}
\begin{small}
\begin{center}
\begin{tabular}{ccccc}
\hline
\multicolumn{1}{c}{Model} &
\multicolumn{1}{c}{$M_\ast$} &
\multicolumn{1}{c}{$R_{\rm e}$} &
\multicolumn{1}{c}{$c$} \\
\multicolumn{1}{c}{} &
\multicolumn{1}{c}{[\Msun]} &
\multicolumn{1}{c}{[pc]} &
\multicolumn{1}{c}{} \\
\hline
C3   & $2.0\times10^6$ & 2.18 & 0.12 \\
C4   & $2.0\times10^5$ & 1.11 & 0.12  \\
C5   & $6.0\times10^5$ & 1.11 & 0.16 \\
\hline
\end{tabular}
\begin{minipage}{8.5cm} {\em Note.} Col.(2): stellar mass.
 Col. (3): half mass radius. Col.(4): concentration parameter.
\end{minipage}
\end{center}
\end{small}
\end{table}

The initial model in run A1 consists of the bulge hosting a nuclear
cluster spheroid (NCS), produced by letting star cluster C3 fall to
the centre starting from a circular orbit at $127$ pc. We use model C5
for the accreted star clusters, starting them on circular orbits at a
distance of $32$ pc from the centre.  Each accretion is allowed to
finish before a new star cluster is inserted.  In total 27 star
clusters, corresponding to $8.1\times$ the NCS's initial mass, are
accreted in 810 Myr.

In model A2 we generated a bare nuclear cluster disc (NCD) model by
adiabatically growing an exponential disc with a scale-length $R_{\rm
  d} = 9.5$ pc and mass of $1\,\times\,10^6$ \Msun\ at the centre of
the bulge model.  We set Toomre-$Q = 1.2$, as described in
\citet{Debattista2000}.  In this model we accrete, over 1.8 Gyr, 50
copies of model C4 sequentially on circular orbits, each starting 63
pc from the centre.  This corresponds to $10\times$ the NCD's initial
mass.

\section{Analysis}
\label{sec:analysis}

\subsection{Pseudo surface photometry and stellar kinematics}
\label{sec:data}

We consider 3 outputs from each run: in model A1 after the accretion
of 10, 20, and 27 star clusters and in model A2 after the accretion of
10, 30, and 50 star clusters.  We start by building an $I$-band image
for each simulation time-step. We model the luminosity distribution of
each galaxy component from its mass distribution by adopting $M/L$
ratios from \citet{Maraston1998, Maraston2005}:

\begin{itemize} 

\item for the NCS of model A1 we assume two different stellar
 components with the same mass:

\begin{enumerate}
\item a young stellar population with an age 1 Gyr, a metallicity
  $\rm [Fe/H]\,=\,-0.4$ dex, and $(M/L)_I\,=\,0.52$ M$_{\odot}$/L$_{\odot}$;

\item an old stellar population with an age 10 Gyr, a metallicity $\rm [Fe/H]
 = -1.4$ dex, and with $(M/L)_I\,=\,2.70$ M$_{\odot}$/L$_{\odot}$;
\end{enumerate}

\item for the  initial NCD of model A2 we adopt a stellar population with
 $(M/L)_I\,=\,0.20$ M$_{\odot}$/L$_{\odot}$ corresponding to an age
 of 70 Myr and a metallicity of $\rm [Fe/H]\,=\,-0.4$ dex;

\item for the host bulge in both model A1 and A2 we consider the same
 population as the old part of the NCS;

\item in each time-step of both model A1 and A2 we consider the same
 mixed population of the NCS for all the star clusters except for the last
 accreted one, for which we adopt the same young population as the
 NCD.
\end{itemize}

These ages and metallicities are chosen to match the properties of the
stellar components observed in the nucleus of NGC~4244
\citep{Seth2006} as in H$11$.  During the evolution time, we consider
that all the components become older, reaching the same mixed
population as the NCS, except for the last accreted star cluster, for
which we adopt the $M/L$ ratio described above. In order to increase
the resolution at the centre of the model, particles in the bulge
model have different masses depending on their total angular momentum;
at larger radii particles have masses as large as $3.9 \times 10^5
\Msun$.  Since these project onto the nucleus, in our analysis of
projected quantities we exclude all bulge particles beyond $200$ pc
from the galaxy centre.

We generate images assuming the Virgo Cluster distance, 16 Mpc.
Therefore, the angular size of 1 arcsec corresponds to 77.6 pc. Images
of the different time-steps are trimmed to match the field of view
($162\,\times\,162$ arcsec$^2$) of the ultraviolet and visible channel
(UVIS) of the Wide Field Camera 3 (WFC3) on board the {\em HST\/},
with a pixel scale of 0.04 arcsec pixel$^{-1}$, and gain and readout
noise set to 1.5 $e^-$ count$^{-1}$ and 3.0 $e^-$ rms, respectively
\citep{Dressler2011}.  Finally, a background level and photon noise
are added to the resulting images to yield a signal-to-noise ratio
similar to that of the {\em HST} images of galaxy nuclei hosting a
NSD.
Given that NSDs are easiest to detect when nearly edge-on, the mock
images to be analysed are generated by adopting a galaxy inclination
of $75^\circ$ which takes into account the detection limit of embedded
discs as a function of their luminosity and inclination
\citep{Rix1990}.

\begin{figure*}
\begin{center}
MODEL A1
\end{center}
\psfig{file=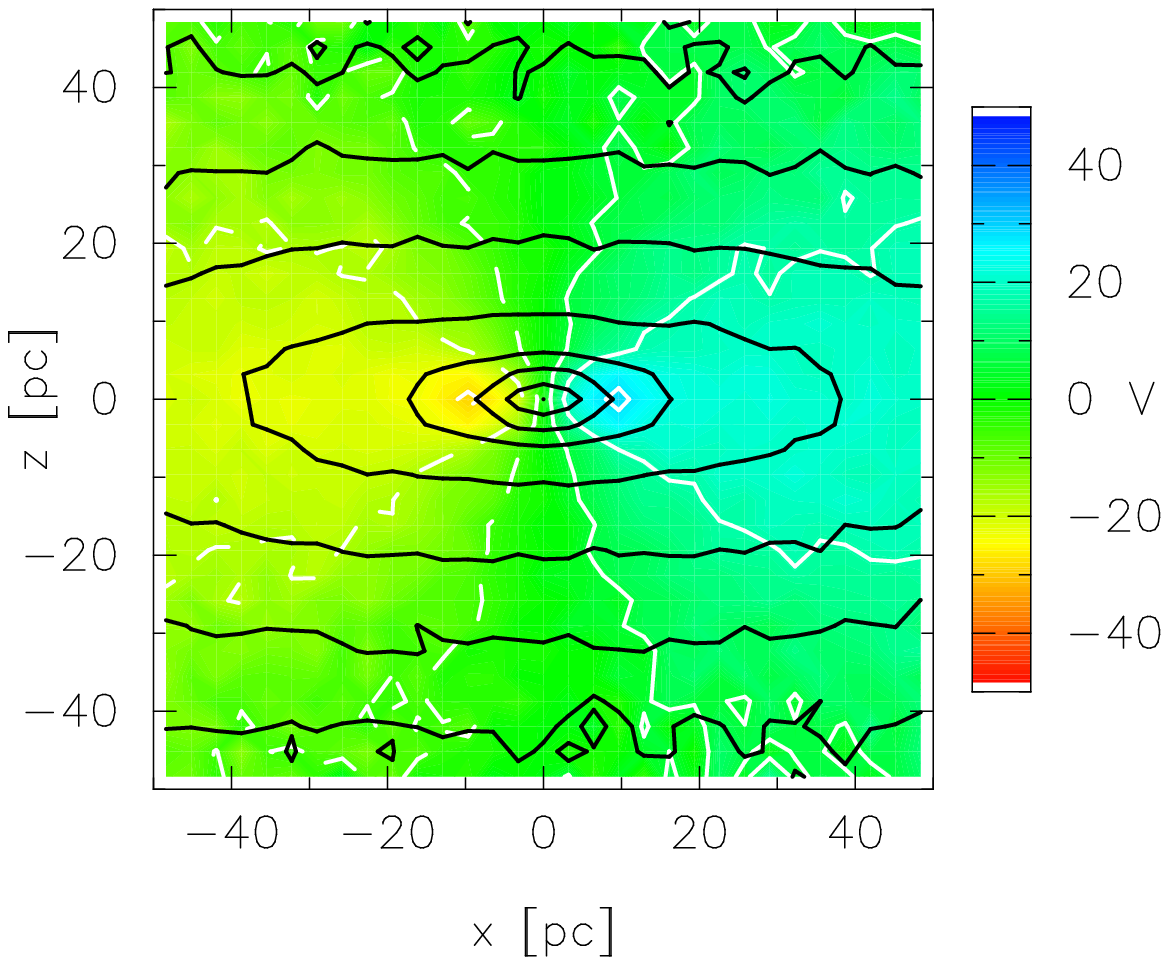,width=4.5cm, angle=-90}
\hspace{0.05cm}\psfig{file=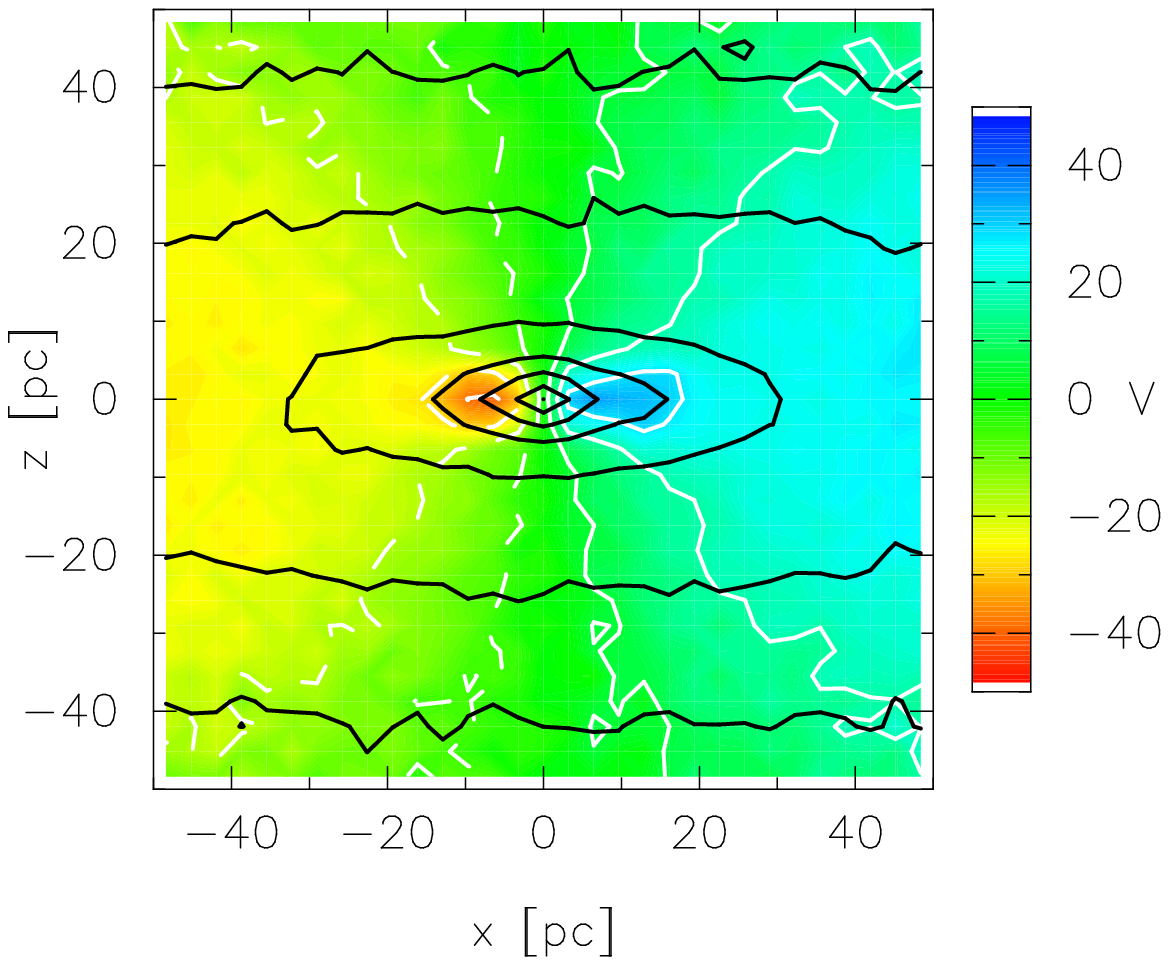,width=4.5cm, angle=-90}
\hspace{0.05cm}\psfig{file=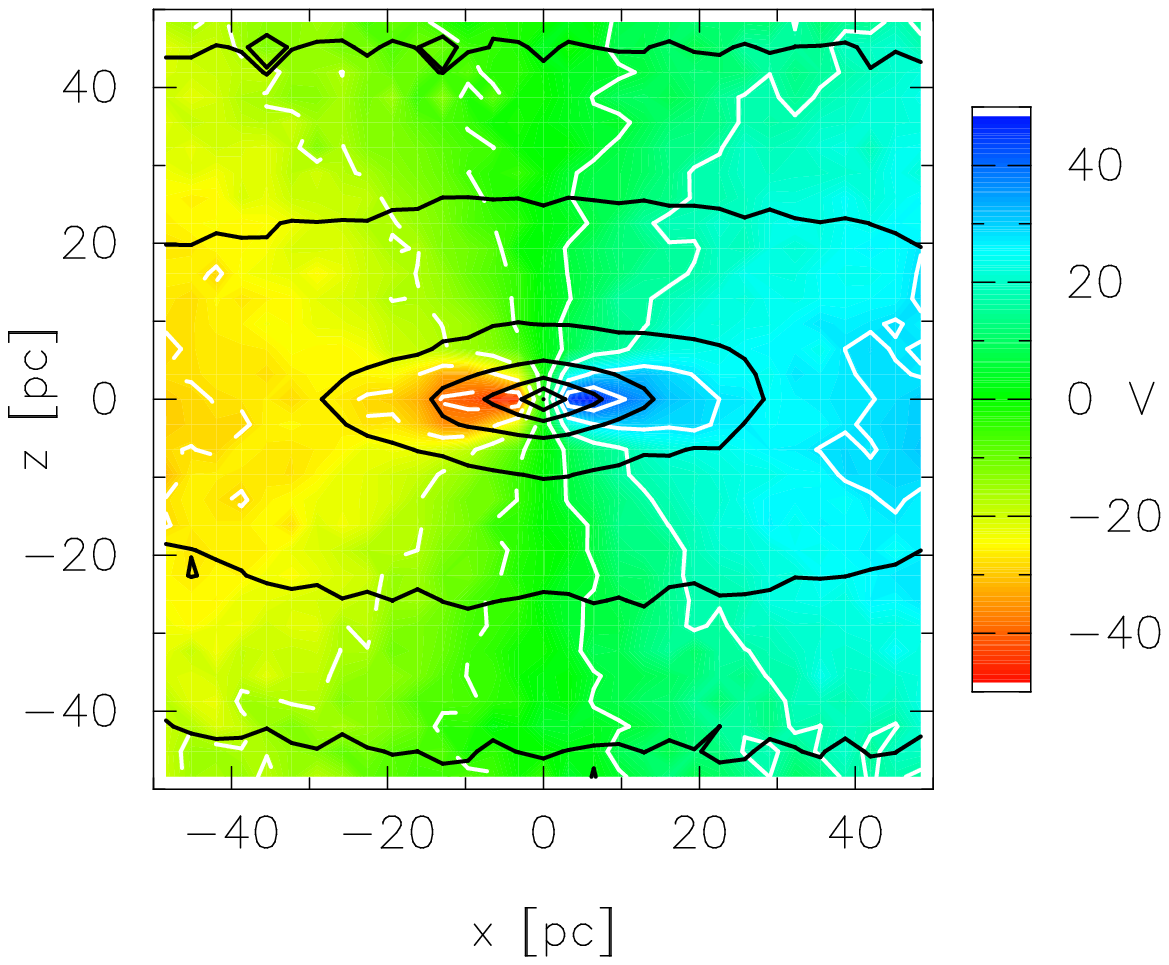,width=4.5cm, angle=-90}

\psfig{file=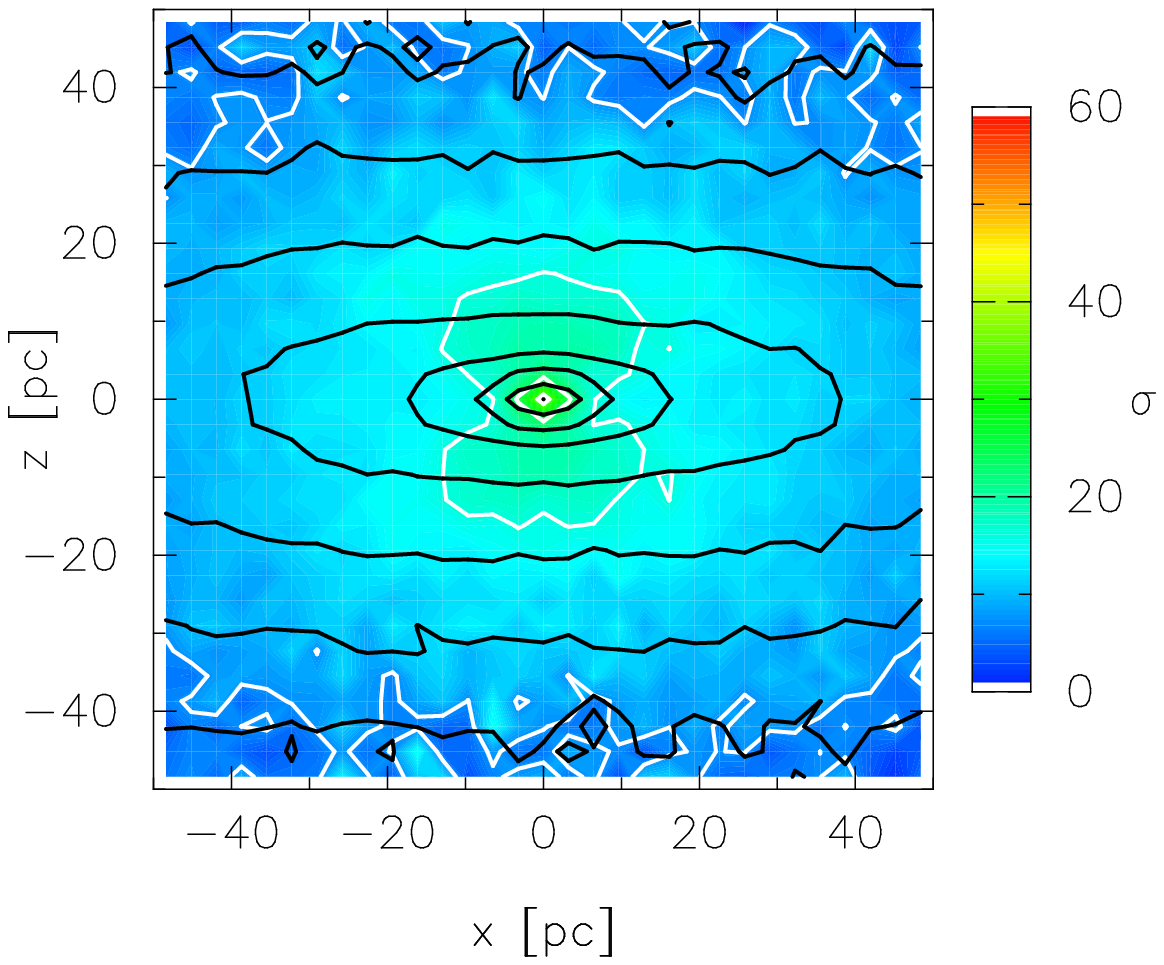,width=4.5cm, angle=-90}
\hspace{0.05cm}\psfig{file=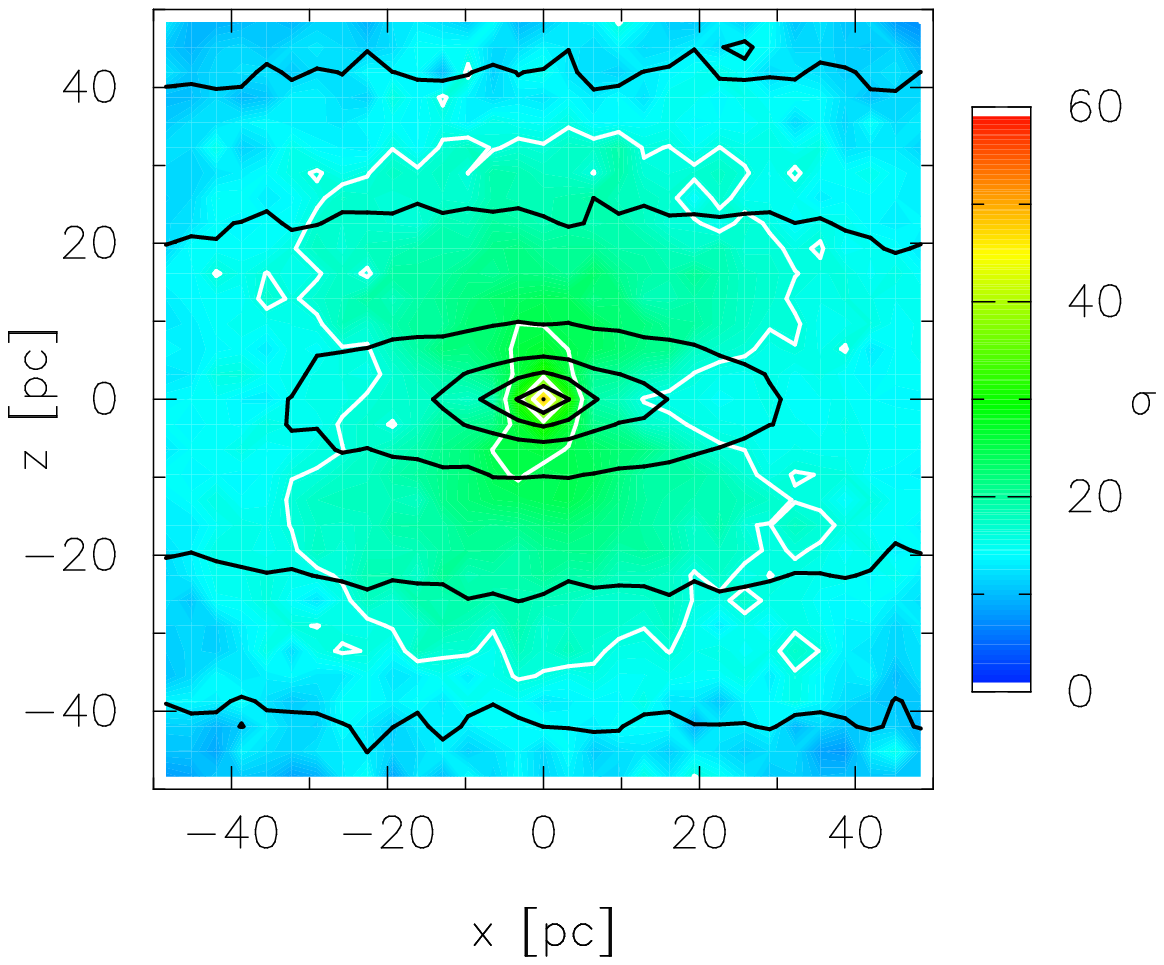,width=4.5cm, angle=-90}
\hspace{0.05cm}\psfig{file=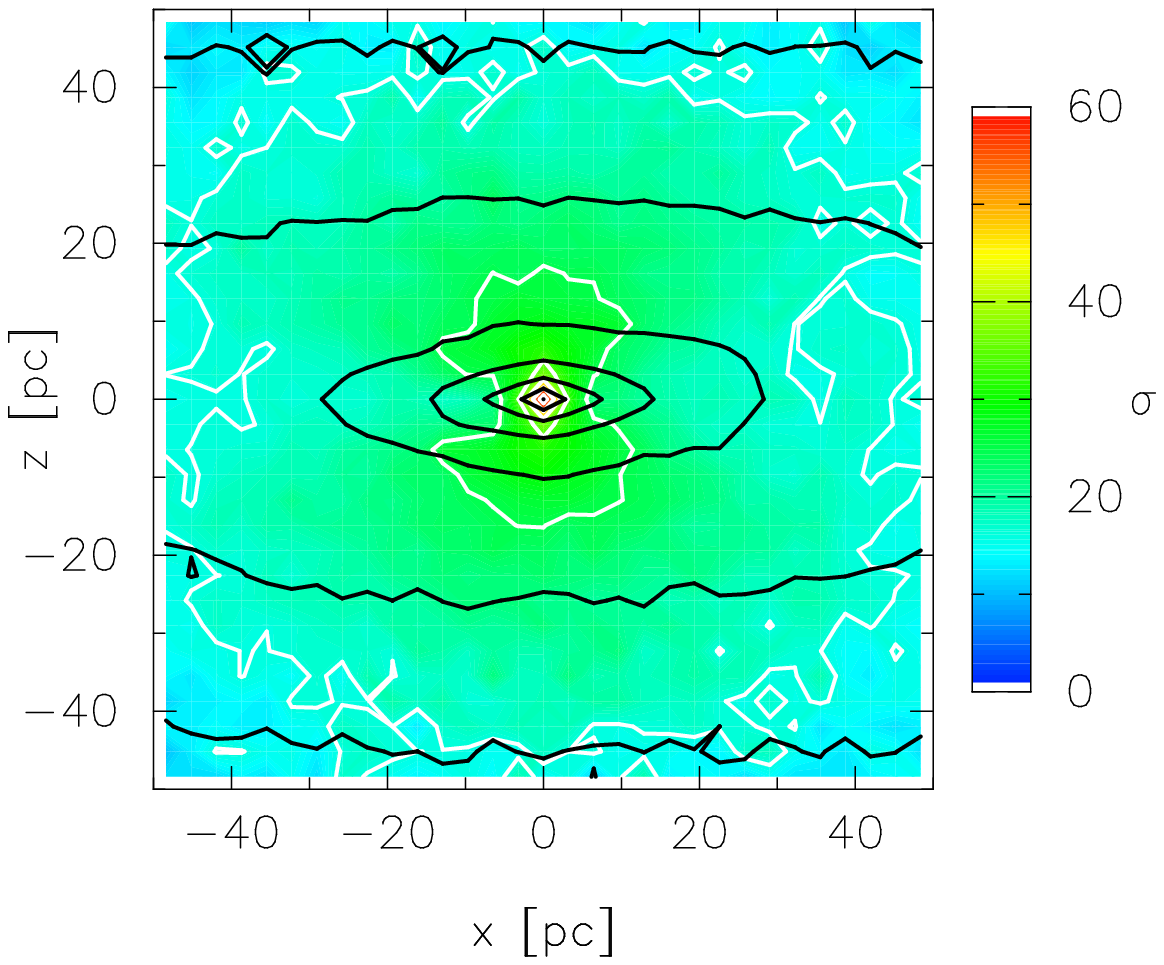,width=4.5cm, angle=-90}

\vspace{0.5cm}

\begin{center}
MODEL A2
\end{center}
\psfig{file=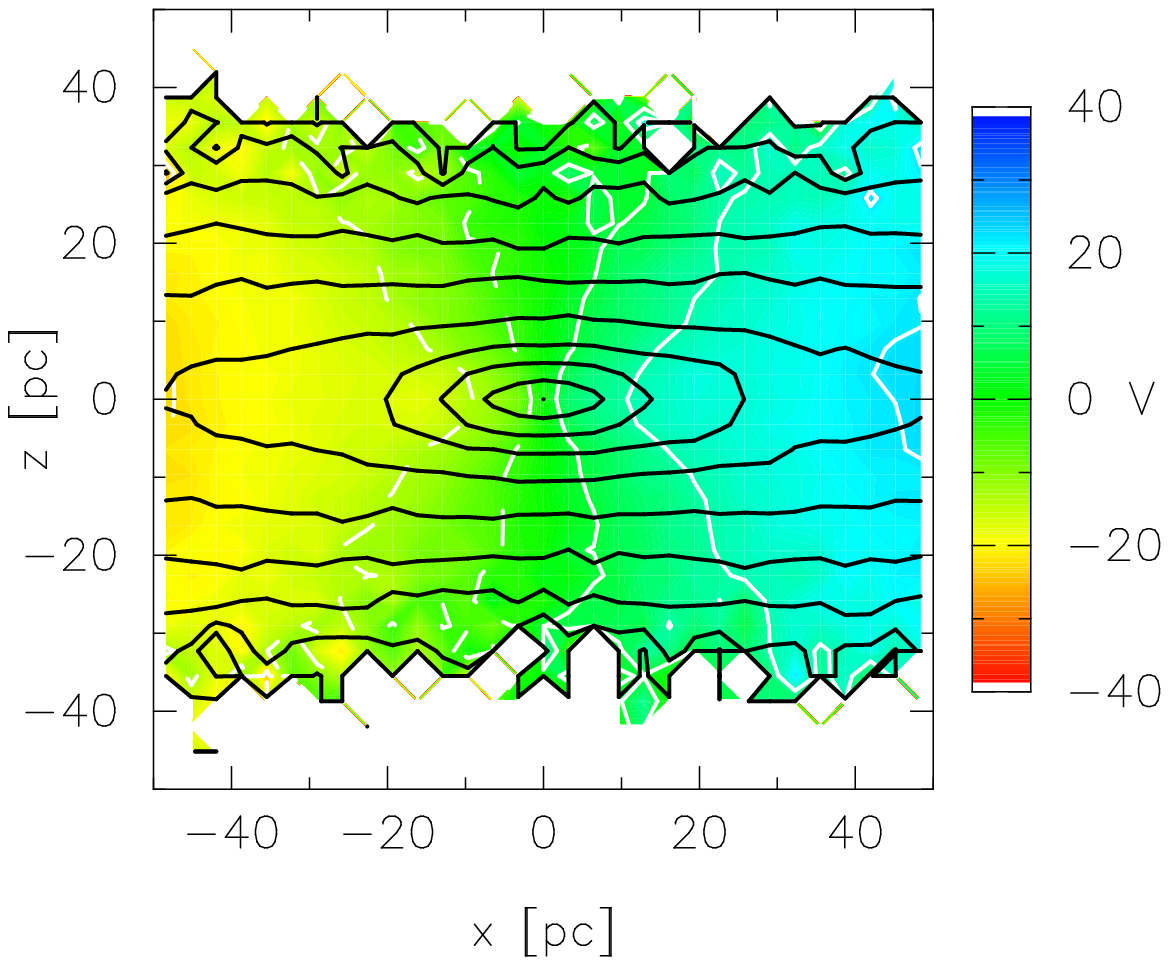,width=4.5cm,angle=-90}
\hspace{0.05cm}\psfig{file=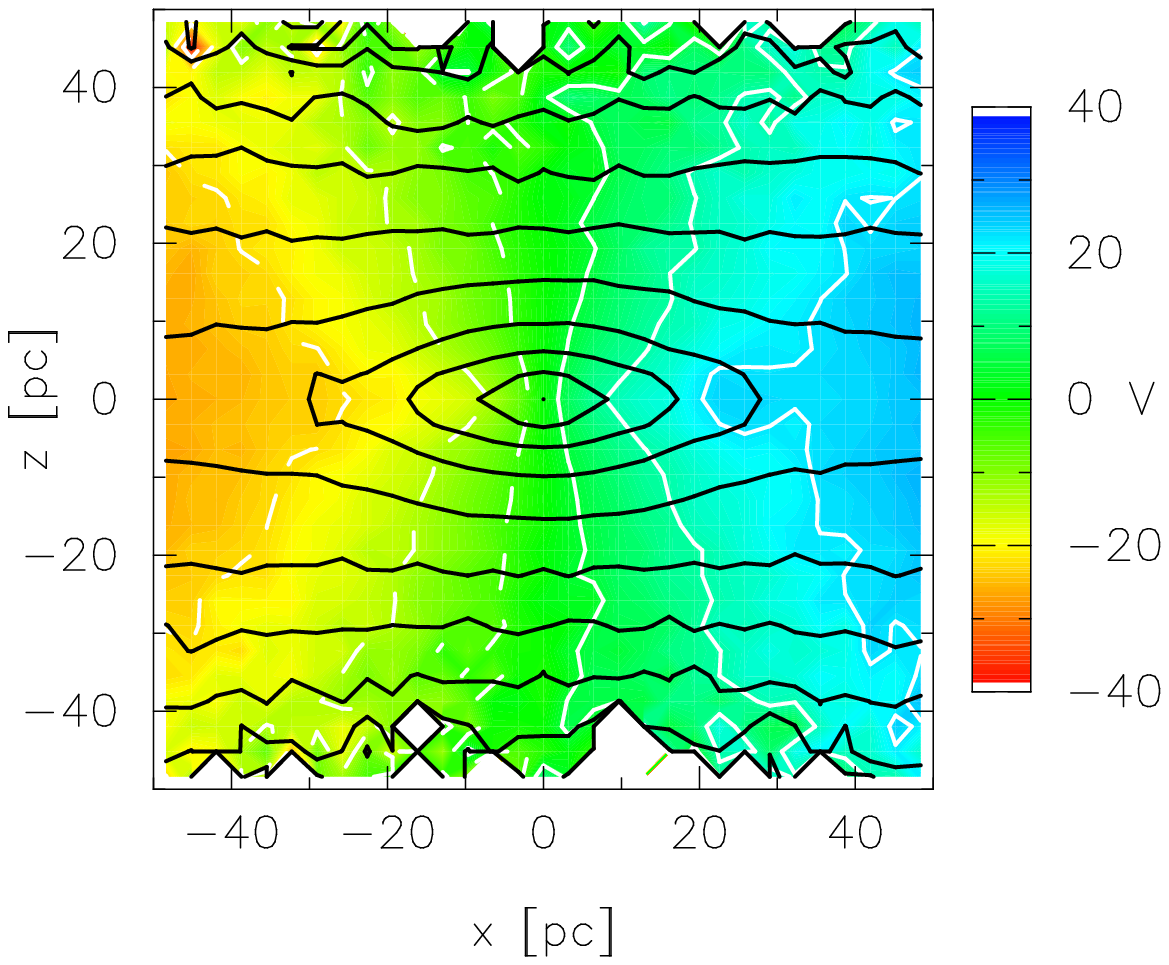,width=4.5cm,angle=-90}
\hspace{0.05cm}\psfig{file=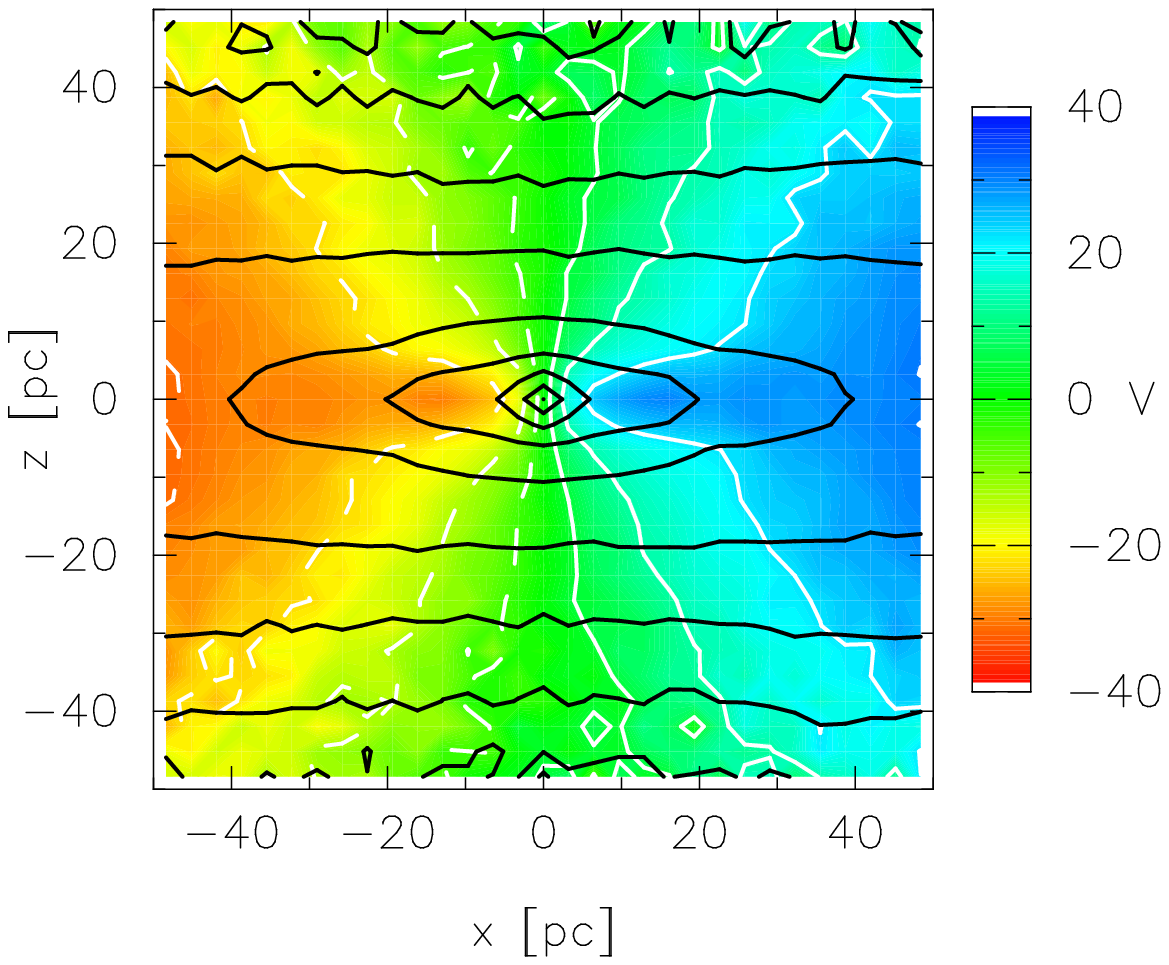,width=4.5cm,angle=-90}

\psfig{file=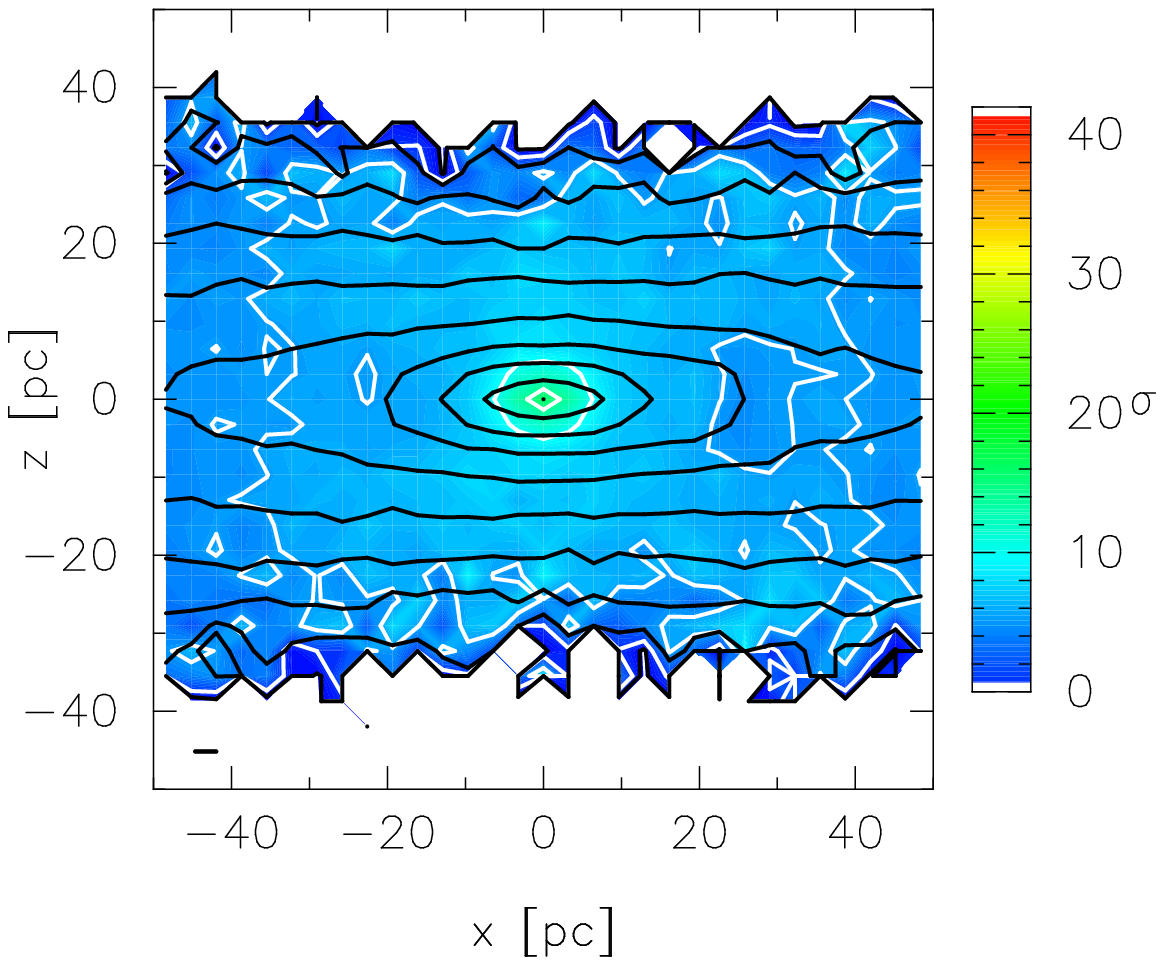,width=4.5cm, angle=-90}
\hspace{0.05cm}\psfig{file=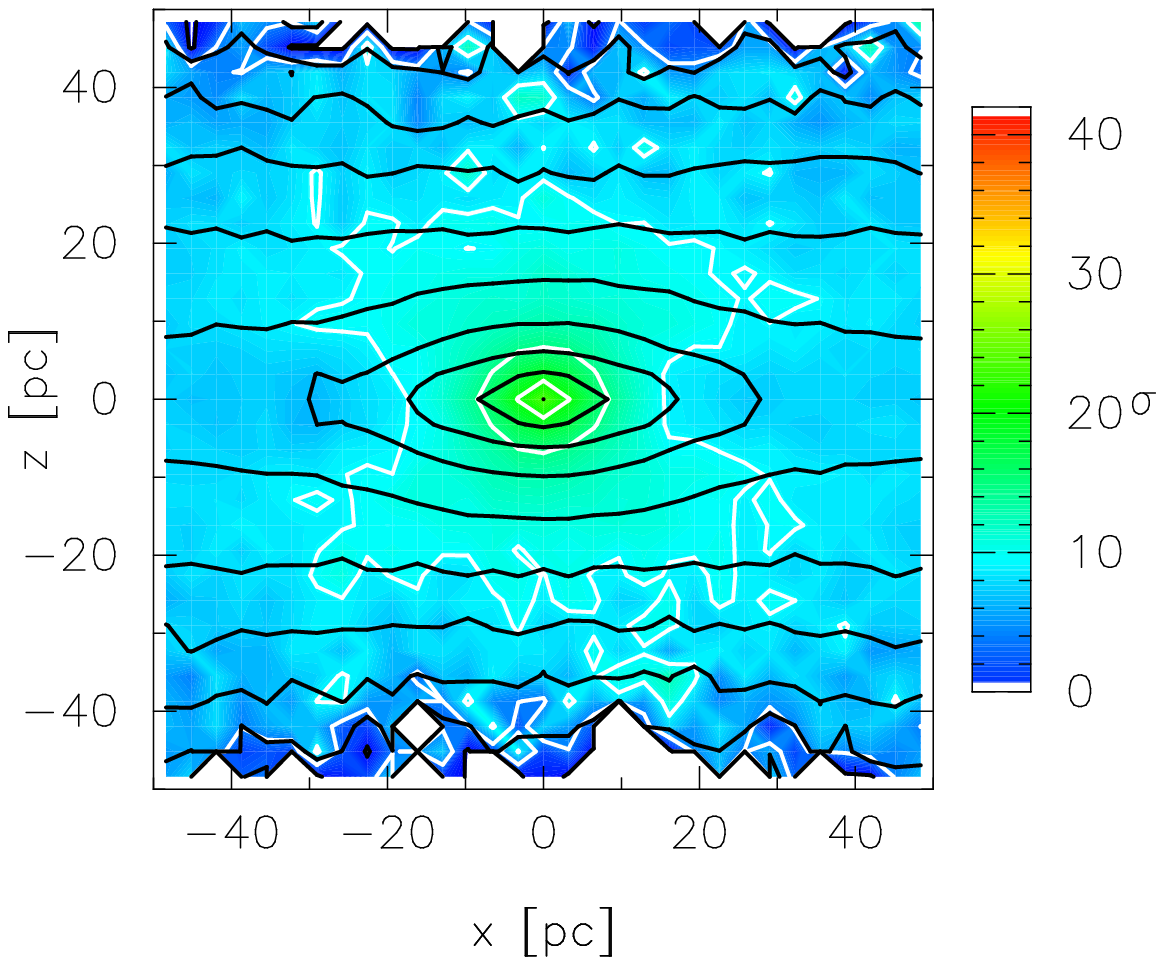,width=4.5cm,angle=-90}
\hspace{0.05cm}\psfig{file=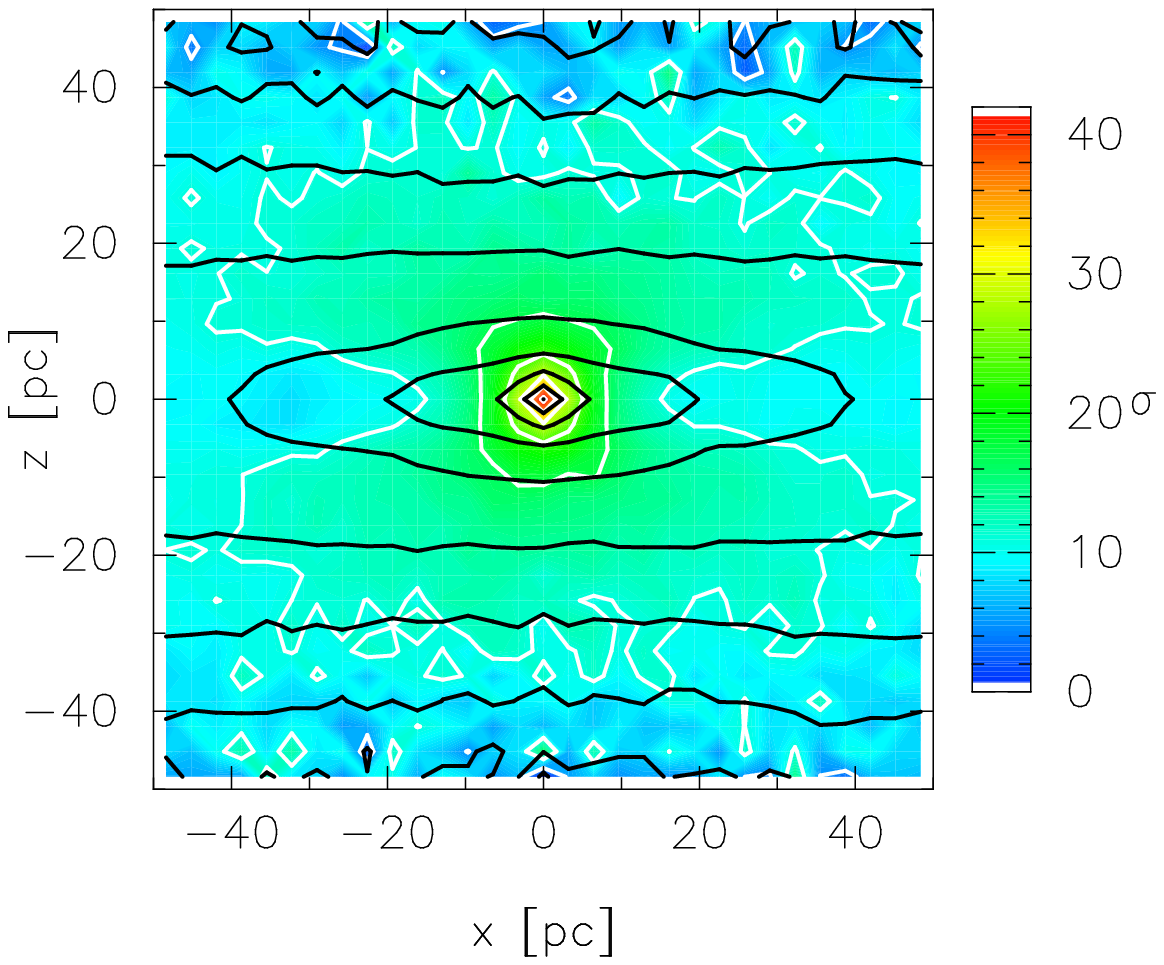,width=4.5cm,angle=-90}

\caption{\small{Top panels: Line-of-sight stellar velocity (top row)
    and velocity dispersion (bottom row) maps for the simulated galaxy
    in model A1 after the accretion of 10 (left-hand panels), 20
    (central panels), and 27 star clusters (right-hand panels). The
    colour code is at the right of each panel. The black contours show
    the galaxy isophotes at a level of $90, 80,\dots,10$ per cent of
    the central surface brightness. The white lines show kinematic
    contours appropriate to each panel. The field of view is
    $0.65\,\times\,0.65$ arcsec$^2$ (corresponding to $50\,\times\,50$
    pc$^2$ at the assumed distance). Bottom panels: As above, but for
    the simulated galaxy in model A2 after the accretion of 10
    (left-hand panels), 30 (central panels), and 50 star clusters
    (right-hand panels).}}
\label{fig:kinematics}
\end{figure*}

The luminosity-weighted kinematics of the stars is derived for each
simulated galaxy as seen edge-on after excluding the contribution of
the bulge to avoid its contamination. The resulting maps of the
line-of-sight velocity and velocity dispersion are shown in
Fig.~\ref{fig:kinematics}.
As pointed out by H$11$, the measured kinematics are dominated by the
nuclear star cluster out to 15 pc from the centre. At larger radii
($r\simeq30$ pc) the amplitude of the stellar rotation
($10\,\la\,V_{\rm max}\,\la\,40$ km~s$^{-1}$) is consistent with
kinematic measurements of actual NSDs
\citep[e.g.,][]{vandenBosch_etal1998, Bertola1999b, Halliday2001}.


\subsection{Detection of the nuclear stellar disc}
\label{sec:unsharp}

\begin{figure*}
\center
\psfig{file=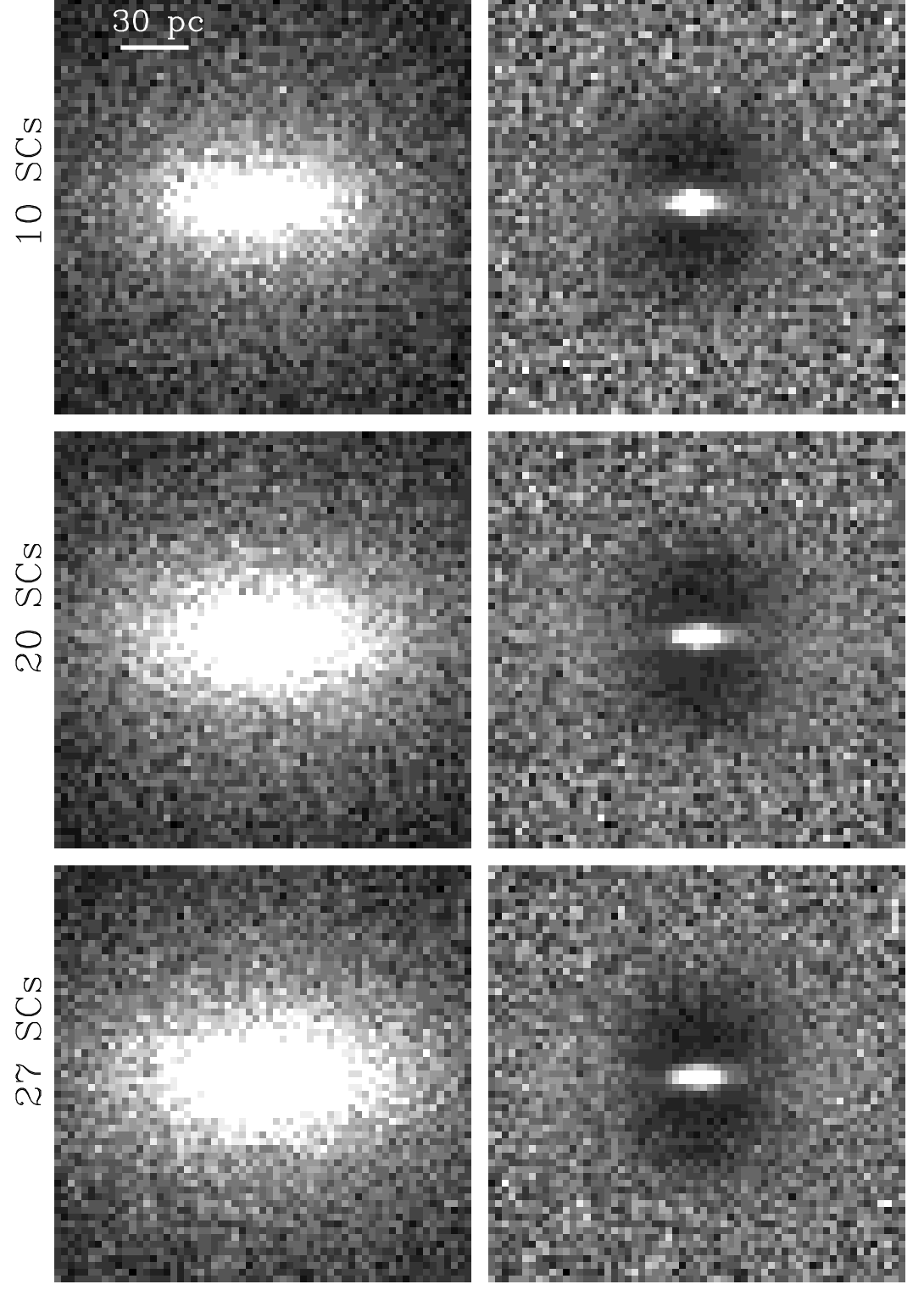,width=8cm}
\psfig{file=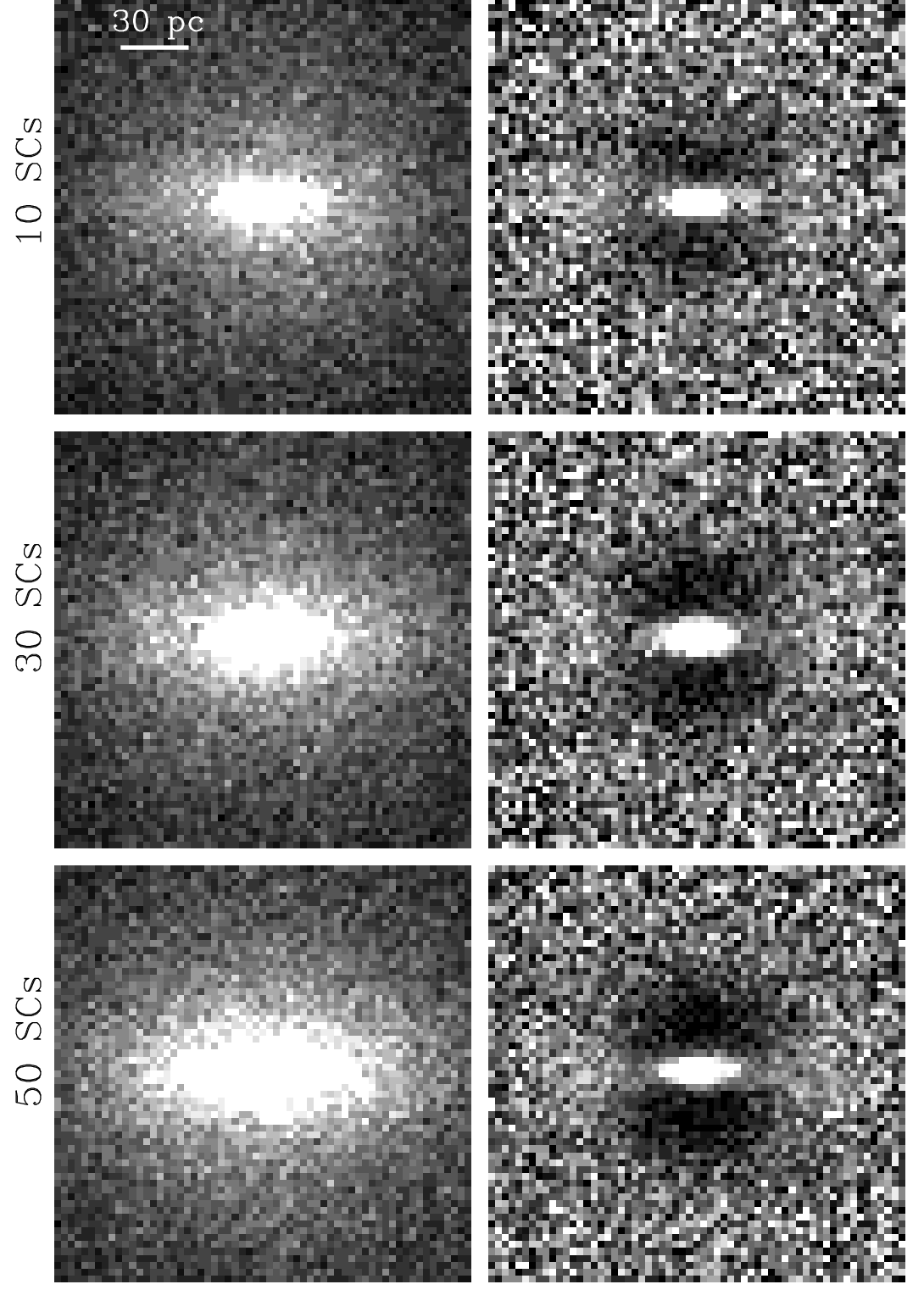,width=8cm}
\caption{\small{Left: mock WFC3/UVIS $I$-band (left-hand panels) and
  unsharp-masked images (right-hand panels) of model A1 after the
  accretion of 10 (top panels), 20 (middle panels), and 27 star
  clusters (bottom panels). The field of view is $2.4\,\times\,2.4$
  arcsec$^2$ (corresponding to $186\,\times\,186$ pc$^2$) and the
  image scale is given in the upper left corner of the first panel.
  The unsharp-masked images were obtained by adopting a smoothing
  Gaussian of $\sigma=10$ pixel corresponding to 0.4 arcsec.  Right:
  As on the left, but for model A2 after the accretion of 10 (top
  panels), 30 (middle panels), and 50 star clusters (bottom panels).
  These images have been obtained at an inclination of $75^\circ$.}}
\label{fig:unsharp}
\end{figure*}

To test for the presence of a NSD in the mock WFC3/UVIS images of the
simulated galaxies, we construct the unsharp-masked image of the
frames as in \citet{Pizzella2002}.  Each image is divided by itself
after convolution with a circular Gaussian of width $\sigma=2,6,10,$
or 20 pixels corresponding to $0.08,0.024,0.4$, or 0.8 arcsec,
respectively (Fig.~\ref{fig:unsharp}).
This procedure enhances any non-circular structure extending over a
spatial scale comparable to the smoothing scale.  Different values of
$\sigma$ are adopted to identify structures of different sizes. For
each model, the location, orientation, and size of the elongated
nuclear component remain similar in all the time-steps.


\citet{Morelli2004} demonstrated that bright, elongated nuclear
structures in unsharp-masked images are not an artifact of the image
processing. Such structures are always observed in galactic nuclei if
an inclined NSD is present. However, the same feature in the
unsharp-masked images of bulges can also be caused by isophotes of
increasing ellipticity inwards. Therefore, to unveil a NSD it is
necessary to perform also a detailed measurement of the
surface-brightness distribution in the nuclear regions.
The isophotes of a bulge are expected to be nearly elliptical, even if
the ellipticity changes with radius. Instead, the presence of an
inclined NSD will result in isophotes with a discy shape, due to the
superposition of the light contribution of the rounder bulge with that
from the more elongated NSD \citep{Scorza1995}. A sharp increase of
the disciness of the innermost isophotes of a galaxy is usually
associated with a strong increase of their ellipticity. The peak of
both ellipticity and disciness in the same radial range is therefore
the full photometric signature of the presence of a NSD embedded in
the bulge.

\begin{figure*}
\center
\psfig{file=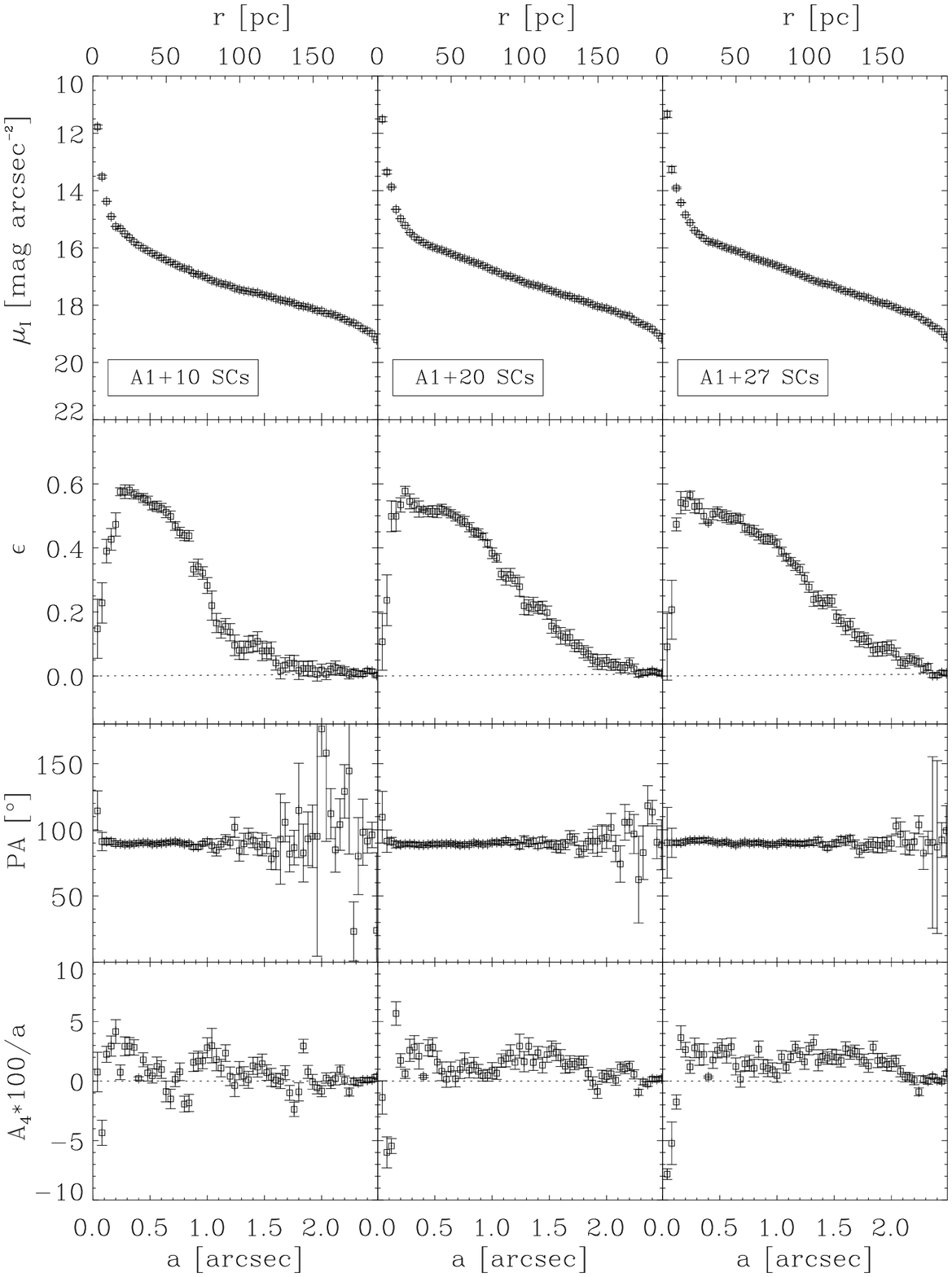,width=0.9\hsize}
\caption{\small{Isophotal parameters of the nuclear region of model A1
    as a function of the isophotal semi-major axis based on the
    analysis of the surface brightness distribution after the
    accretion of 10 (left-hand panels), 20 (central panels), and 27
    (right-hand panels) star clusters.  From top to bottom: Radial
    profiles of the surface-brightness, ellipticity, position angle,
    and fourth cosine Fourier coefficient. }}
\label{fig:ellipse_A1}
\end{figure*}

\begin{figure*}
\center
\psfig{file=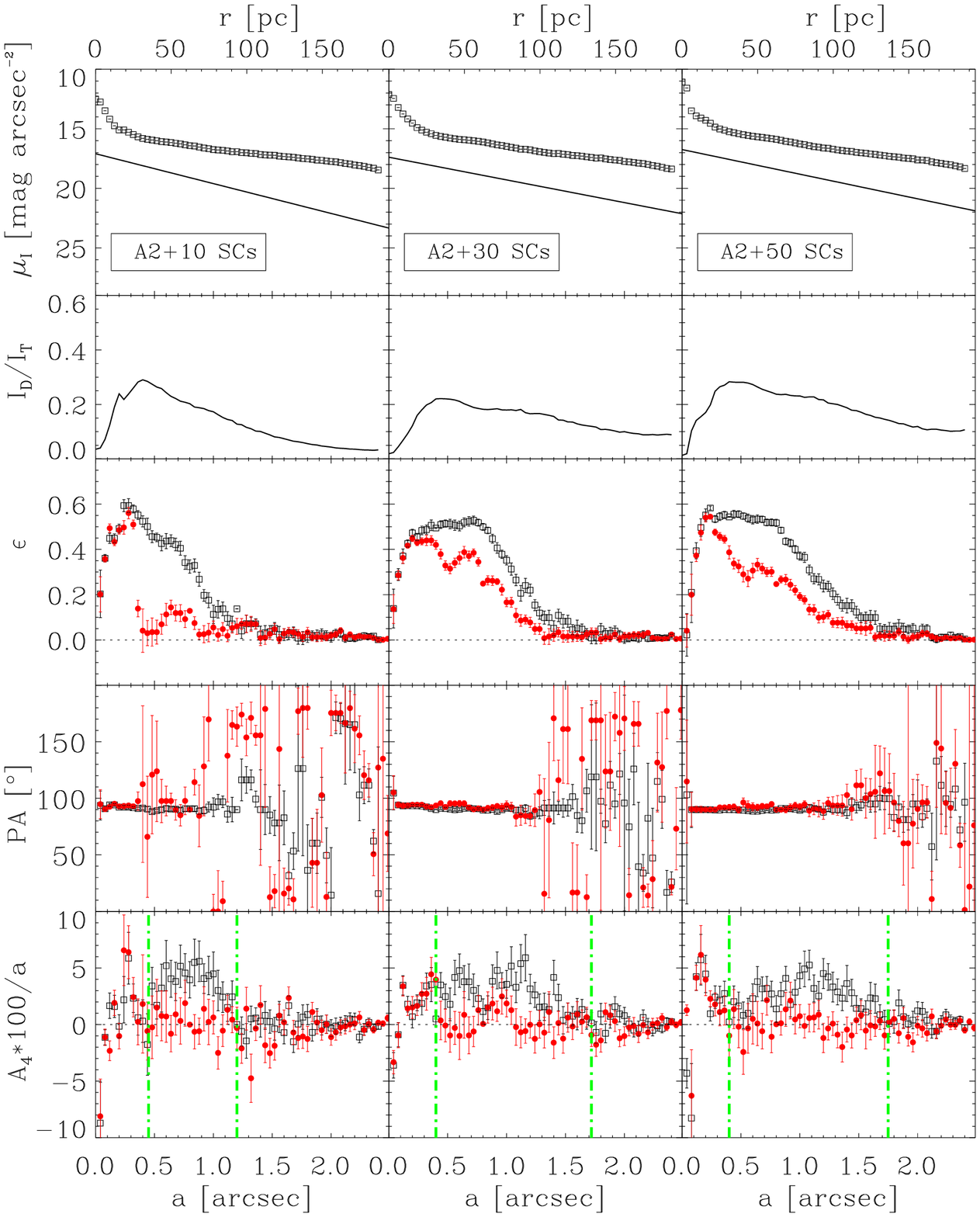,width=0.9\hsize }
\caption{\small{Isophotal parameters of the nuclear region of model A2
    as a function of the isophotal semi-major axis based on the
    analysis of the surface brightness distribution measured in the
    mock images obtained after the accretion of 10 (left-hand panels),
    30 (central panels), and 50 (right-hand panels) star clusters.
    From top to bottom: surface-brightness radial profiles of the
    galaxy (open black squares) and NSD (solid line), radial profiles
    of the NSD-to-total surface-brightness ratio, radial profiles of
    the galaxy ellipticity, position angle and fourth cosine Fourier
    coefficient before (open black squares) and after (filled red
    circles) the subtraction of the best-fitting model for the
    NSD. The vertical (green) dot-dashed lines mark the radial range
    within which the light distribution is characterised by discy
    isophotes ($A_4>0$) due to the presence of the NSD.}}
\label{fig:ellipse_A2}
\end{figure*}

We fit isophotes to the model images using the {\sc
IRAF}\footnote{Imaging Reduction and Analysis Facilities (IRAF) is
distributed by the National Optical Astronomy Observatories which
are operated by the Association of Universities for Research in
Astronomy (AURA) under cooperative agreement with the National
Science Foundation.} task {\sc ELLIPSE} \citep{Jedrzejewski1987}.
Isophotes are fitted with ellipses, allowing their centres to vary in
order to look for asymmetries in the light distribution. Within the
errors of the fits we find no evidence of variations in the fitted
centres. The ellipse fitting is then repeated with the ellipse centres
fixed. The resulting azimuthally-averaged surface brightness,
ellipticity, position angle, and fourth cosine Fourier coefficient
($A_4$) profiles are presented in Figs.  \ref{fig:ellipse_A1} and
\ref{fig:ellipse_A2} for models A1 and A2, respectively.  Positive
values of the $A_4$ Fourier coefficient are characteristic of discy
isophotes \citep{Bender1988}.

The radial profile of ellipticity peaks at about 0.2 arcsec from the
centre ($\epsilon_{\rm max} \simeq 0.6$) in all the mock images of
model A1. Round isophotes ($\epsilon \simeq 0$) are observed for radii
larger than 2 arcsec. Within the innermost 2 arcsec there is no change
in the $A_4$ Fourier coefficient, which is nearly constant and
slightly positive ($A_4 \simeq 0.01$ after the accretion of 10 and 20
star clusters, increasing to $A_4 \simeq 0.02$ after the accretion of
27 star clusters).  Thus model A1 never develops the photometric
signature of observed NSDs.

In model A2, instead, both the ellipticity and the $A_4$ Fourier
coefficient show a sharp increase within 1.5 arcsec from the centre in
all the time-steps. The maximum observed values of ellipticity and
disciness are $\epsilon_{\rm max} \simeq 0.6$ and $A_{4, {\rm max}}
\simeq 0.05$, respectively. We interpret these photometric features as
the signature of an embedded NSD in the bulge of model A2.  Indeed the
same photometric features are observed in galactic nuclei hosting a
NSD \citep[e.g.,][]{Scorza1998, Krajnovic2004, Morelli2010}.

\subsection{Photometric decomposition}
\label{sec:decomposition}

We first measure the PA of the major axis of the NSD  found in
  model A2 in order to check
whether we can successfully recover the orientation of the NSD.  We
assume the images are oriented with the North up and East left and
analyse them after unsharp-masking. The images are rotated from
$0^\circ$ to $179\fdg8$ in steps of $0\fdg2$ and we measure the flux
within a horizontal strip of $51\,\times\,9$ pixels crossing the
galaxy centre. The maximum flux is measured when the strip is aligned
with the NSD major axis. In this way we find $\rm
PA\,=\,90^\circ\pm2^\circ$, consistent with the input line-of-nodes of
the system.

We then apply a photometric decomposition of the surface brightness
based on the Scorza-Bender method \citep{Scorza1995}, as implemented by
\citet{Morelli2004}, to the images of model A2.  The photometric
decomposition, which is performed independently for each time-step, is
based on the assumption that both the host bulge and the inclined NSD
are each characterised by elliptical isophotes with constant
ellipticity. The method consists of an iterative subtraction of
different models of an infinitesimally thin exponential disc.  The
disc has central-surface brightness $I_0$, scale-length $h$, and apparent
axial ratio $q$. We assume $\rm PA\,=\,90^\circ$ for the position
angle of the disc major axis.

For real images it is crucial to first deconvolve the
surface-brightness distribution from the effects of the {\em HST}
point spread function (PSF) in order to properly derive the
photometric parameters of the NSDs \citep[e.g.,][]{Pizzella2002,
 Morelli2010, vandenBosch_Emsellem1998, Krajnovic2004}.  We do not
apply any deconvolution to our mock WFC3/UVIS images since they have
not been convolved with any PSF to begin with.

The remaining NSD parameters are adjusted until the departures from
perfect ellipses are minimised (i.e., $A_4\simeq0$ over all the
observed radial range). For each disc model, first we obtain the
disc-free image of the galaxy by subtracting the disc model from the
galaxy image. Then, we perform an isophotal analysis of the disc-free
image using {\sc ELLIPSE}. We calculate:
\begin{equation}
\chi^2=\sum_{i=1}^N{\frac{A^2_{\rm{4,disc-free}}(i)}{\sigma^2 (i)}}
\end{equation}
where $A_{\rm{4,disc-free}}(i)$ is the value of the $A_4$ Fourier
coefficient measured for the $i-$th isophote in the disc-free image,
and $N$ is the total number of fitted isophotes. We assume $\sigma(i)$
= 0.01 as a typical error on $A_{\rm{4,disc-free}}$ for all the
isophotes in the region of the NSD. The NSD region is bracketed by two
vertical lines in Fig.~\ref{fig:ellipse_A2}.  The minimum value of
$\chi^2$ corresponds to the best-fitting model of the NSD. We
determine $\Delta \chi^2 \equiv \chi^2 - \chi^2_{\rm min}$ and derive
its confidence levels under the assumption that the errors are
normally distributed. The resulting contour plots of $\chi^2$ allow us
to derive the best-fitting values of $I_0$, $h$, and $q$, and their
$3\sigma$ uncertainties.
The inclination is calculated as $i = \arccos{q}$.
The resulting values of the observed central surface-brightness,
scale-length, inclination, and total luminosity of the NSD in the
$I$-band for the different time-steps of model A2 are listed in
Table~\ref{tab:photometry}.

\begin{table}
\caption{\small{Photometric parameters of the nuclear stellar discs
   derived from the photometric decomposition of the mock images of
   model A2.}}  
\setlength{\tabcolsep}{1.4mm}
\label{tab:photometry}
\begin{small}
\begin{center}
\begin{tabular}{cccccc}
\hline
\noalign{\smallskip}
\multicolumn{1}{c}{$N_{\rm SC}$} &
\multicolumn{1}{c}{Fit} &
\multicolumn{1}{c}{$\mu_{0,I}$} &
\multicolumn{1}{c}{$h$} &
\multicolumn{1}{c}{$i$} &
\multicolumn{1}{c}{$L_{{\rm T},I}$} \\
\multicolumn{1}{c}{} &
\multicolumn{1}{c}{} &
\multicolumn{1}{c}{[mag arcsec$^{-2}$]} &
\multicolumn{1}{c}{[arcsec]} &
\multicolumn{1}{c}{[$^{\circ}$]} &
\multicolumn{1}{c}{[$10^6$ L$_{\odot}$]} \\
\hline
10 & SB & $17.09^{+0.03}_{-0.11}$ & $0.43^{+0.02}_{-0.04}$ & $75.5^{+1.5}_{-1.9}$ & $4.7^{+0.1}_{-0.5}$\\
10 & EF & $16.92\pm0.20$         & $0.40\pm0.04$       & $77.9\pm1.4$       & $3.8^{+0.5}_{-1.0}$\\
30 & SB & $17.39^{+0.31}_{-0.14}$ & $0.54^{+0.08}_{-0.12}$ & $78.3^{+2.3}_{-3.3}$ & $4.6^{+1.2}_{-0.6}$\\
30 & EF & $17.11\pm0.20$         & $0.52\pm0.04$       & $76.7\pm0.8$       & $6.8^{+1.3}_{-1.2}$\\
50 & SB & $16.94^{+0.19}_{-0.15}$ & $0.52^{+0.07}_{-0.08}$ & $78.1^{+3.1}_{-2.8}$ & $6.5\pm1.0$\\
50 & EF & $16.85\pm0.10$         & $0.54\pm0.04$       & $77.3\pm0.6$       & $7.6\pm0.7$\\
\hline
\end{tabular}
\begin{minipage}{8.5cm} {\em Note.} Col.(1): number of accreted star clusters.
 Col.(2): photometric decomposition method: SB = Scorza-Bender method
 on the mock images including the light contribution of the bulge, EF
 = exponential fit on the mock images excluding the light contribution
 of the bulge.  Col.(3): observed central surface-brightness. Col.(4):
 scale-length. Col.(5): inclination.  Col.(6): total $I$-band
 luminosity.
\end{minipage}
\end{center}
\end{small}
\end{table}

We test our decompositions by an independent analysis of an image
which excludes all bulge particles. We build an $I$-band image of each
time-step with the same assumptions as in Sect.~\ref{sec:data}
including only the light contribution of the bare NCD and accreted
star clusters, which are expected to be the building blocks of the
NSD. The resulting surface-brightness distribution is modelled with a
S\'ersic component and an exponential disc by applying the
two-dimensional fitting algorithm {\sc GALFIT} \citep{Peng2002}. The
S\'ersic component accounts for the central bright structure which
dominates the light distribution in the innermost 0.2 arcsec
(Fig.~\ref{fig:ellipse_A2}).  The best-fitting parameters of the
exponential disc are in agreement within errors with the NSD
parameters we derived with the Scorza-Bender method, as shown in
Table~\ref{tab:photometry}.

We also applied the Scorza-Bender method to model A1. In this case it
failed to find a reasonable NSD, with the best model having a
scale-length $h=0$ pc.

\subsection{Rotation parameter}
\label{sec:rotation}

We compute the rotation parameter as:
\begin{equation}
\frac{V}{\sigma} = 
\sqrt{\frac{\langle V^2\rangle}{\langle \sigma^2\rangle}} =
\sqrt{\frac{\Sigma_{n=1}^N F_n V_n^2}{\Sigma_{n=1}^N F_n \sigma_n^2}},
\end{equation}
and the mean ellipticity as:
\begin{equation}
\epsilon = 1-q =
1 - \sqrt{\frac{\langle y^2\rangle}{\langle x^2\rangle}} =
1 - \sqrt{\frac{\Sigma_{n=1}^N F_n y_n^2}{\Sigma_{n=1}^N F_n x_n^2}},
\end{equation}
where $q$ is the axis ratio and, following H$11$, $V_n$, $\sigma_n$,
and $F_n$ are the line-of-sight velocity, velocity dispersion, and
flux of the pixel at $(x_n,y_n)$, respectively.  The origin of the
Cartesian coordinates $(x,y)$ is at the centre of the galaxy, with the
$x$-axis aligned with the line-of-nodes. We consider only the $N$
pixels between 15 and 40 pc from the galaxy centre to calculate the
luminosity-weighted values of $\langle V^2 \rangle$, $\langle \sigma^2
\rangle$, $\langle x^2 \rangle$ and $\langle y^2 \rangle$ in order to
exclude the contribution of the pre-existing nuclear structure (i.e.,
the NCS in model A1 and the NCD in model A2). There is no need to
correct for inclination, since the kinematic maps are built with the
simulated galaxies edge-on.  We also derive the ratio,
$(V/\sigma)^\ast$, of the measured $V/\sigma$ to the value predicted
for an edge-on, isotropic oblate system flatted by rotation with the
same intrinsic ellipticity as the elongated structure
\citep[see][]{Kormendy1982}.

The resulting values of the luminosity-weighted velocity, velocity
dispersion, and ellipticity together with the rotation parameter
$(V/\sigma)$ and $(V/\sigma)^\ast$ for all time-steps are listed in
Table~\ref{tab:kinematics}. The axis ratio of the elongated structure
measured in the nucleus of model A1 is $q\simeq0.4$. The stellar
kinematics and ellipticity are therefore consistent with an
anisotropic rotator with $(V/\sigma)^\ast\simeq0.6$.  The elongated
structure of model A2 ($q\simeq0.3$) instead rotates almost as fast as
an isotropic oblate system, $(V/\sigma)^\ast\simeq0.8$.
In both models, $V$ and $\sigma$ increase with the number of accreted
star clusters, although the difference between the values measured in the first
and last time-step is smaller than the typical $1\sigma$ error ($5-10$
km~s$^{-1}$) on the stellar kinematics measured in galactic nuclei
\citep[e.g.,][]{Emsellem2004}.

\begin{table}
\caption{\small{Kinematic parameters of the nuclear elongated
   structure observed in models A1 and A2.}}
\setlength{\tabcolsep}{1.4mm}
\label{tab:kinematics}
\begin{small}
\begin{center}
\begin{tabular}{cccccccc}
\hline
\multicolumn{1}{c}{Model} &
\multicolumn{1}{c}{$N_{\rm SC}$} &
\multicolumn{1}{c}{$\frac{M_{\rm SC}}{M_{\rm in}}$} &
\multicolumn{1}{c}{$V$} &
\multicolumn{1}{c}{$\sigma$} &
\multicolumn{1}{c}{$\frac{V}{\sigma}$} &
\multicolumn{1}{c}{$\epsilon$} &
\multicolumn{1}{c}{$\left(\frac{V}{\sigma}\right)^\ast$} \\
\multicolumn{1}{c}{} &
\multicolumn{1}{c}{} &
\multicolumn{1}{c}{} &
\multicolumn{1}{c}{[km s$^{-1}$]} &
\multicolumn{1}{c}{[km s$^{-1}$]} &
\multicolumn{1}{c}{} &
\multicolumn{1}{c}{} &
\multicolumn{1}{c}{} \\
\hline

A1 & 10 & 3   & 15.1 & 15.7 & 0.95 & 0.65 & 0.70 \\
A1 & 20 & 6   & 15.7 & 19.4 & 0.80 & 0.62 & 0.63 \\
A1 & 27 & 8.1 & 16.7 & 21.8 & 0.77 & 0.60 & 0.63\\
\hline               
A2 & 10 & 2 & 12.6 &  7.6 & 1.66 & 0.75 & 0.96\\
A2 & 30 & 6 & 14.2 & 11.0 & 1.29 & 0.68 & 0.82\\
A2 & 50 &10 & 18.5 & 13.8 & 1.34 & 0.72 & 0.84\\
\hline
\end{tabular}
\begin{minipage}{8.5cm} {\em Note.} Col. (1): model number.  Col. (2):
 number of accreted star clusters. Col. (3): ratio of the accreted
 mass to initial mass of the nuclear star cluster. Cols. (4) and (5):
 luminosity-weighted velocity and velocity dispersion excluding the
 contribution of the central pre-existing structure. Col. (6):
 rotation parameter. Col.  (7): luminosity-weighted
 ellipticity. Col. (8): ratio of $V/\sigma$ to the value predicted for
 an edge-on oblate isotropic rotator with an intrinsic ellipticity
 given in Col. (7) and flattened by rotation.
\end{minipage}
\end{center}
\end{small}
\end{table}

\section{Discussion}
\label{sec:discussion}

From the analysis of the unsharp masked image, the nucleus in model A1
shows an elongated structure (Fig.~\ref{fig:unsharp}, left-hand
panels). Its radial extent increases with the number of accreted star
clusters as seen from the increase of the radial range where
significantly elongated isophotes are measured
(Fig.~\ref{fig:ellipse_A1}). Indeed, $\epsilon > 0$ out to 1.1, 2.1,
and 2.4 arcsec from the centre after the accretion of 10, 20, and 27
star clusters. In all the time-steps the ellipticity peaks at
$\epsilon_{\rm max} \simeq 0.6$ at about 0.2 arcsec (15 pc), where the
NCS dominates, as shown in the analysis of H$11$, and it gently
decreases outwards. The absence of a sharp peak in the $A_4$ radial
profile at the location where NSDs reside is the main photometric
reason for concluding that no NSD is present, even if the unsharp mask
technique shows an elongated structure.  Such feature can also be
caused by isophotes of increasing ellipticity inwards. Therefore
unveiling a NSD requires performing a detailed measurement of the
surface-brightness distribution in the nuclear regions. In addition,
the rotation parameter and ellipticity measured from the stellar
kinematics are consistent with an anisotropic rotation
(Table~\ref{tab:kinematics}). Therefore, we conclude that the
elongated nuclear structure observed in model A1 is not a NSD, in
spite of the fact that the total accreted mass, $1.6\times10^7$
M$_\odot$, is comparable to that of observed NSDs
\citep[e.g.,][]{Morelli2004}.

On the other hand, the photometric and kinematic properties of the
nuclear structure seen in model A2 (Fig.~\ref{fig:unsharp}, right-hand
panels) are reminiscent of a NSD.  The rising ellipticity is
associated with discy isophotes over a radial range which steadily
increases with the number of accreted star clusters
(Fig.~\ref{fig:ellipse_A2}).  The ellipticity peaks at a similar value
as model A1 ($\epsilon_{\rm max} \simeq 0.6$) although it is $\epsilon
\simeq 0.5$ over a wider radial range ($0.2-0.8$ arcsec). At larger
radii, the isophotes quickly become round. The elongated structure
rotates as fast as an oblate isotropic spheroid
(Table~\ref{tab:kinematics}). Its surface-brightness distribution can
be modelled as an inclined exponential disc with a maximum local light
contribution to the total surface brightness of about $30$ per cent at
0.4 arcsec from the centre. The NSD parameters measured in the three
time-steps are consistent with each other within the errors.  Their
mean values from the Scorza-Bender method are $\mu_{0,I}\,=\,17.14$
mag arcsec$^{-2}$, $h\,=\,0.50$ arcsec (38.8 pc), and $i\,=\,77\fdg3$
corresponding to $L_{{\rm T},I}\,=\,5.3\times10^6$ L$_\odot$. The
inclination and PA of the line-of-nodes of the NSD and its host bulge
are the same within the errors. The total accreted mass is
$1.0\times10^7$ M$_\odot$ and it is consistent with that of observed
NSDs.

\begin{figure}
\center
\psfig{file=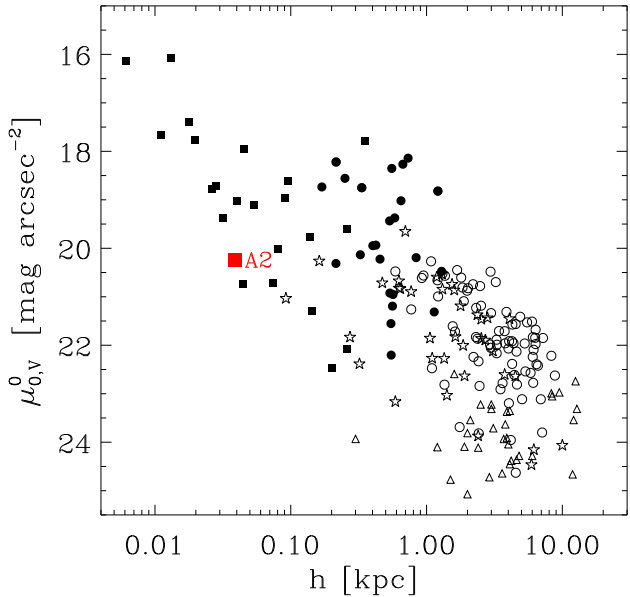,width=8.5cm,bbllx=100, bblly=370, bburx=590, bbury=840}
\caption{\small{The face-on central surface-brightness as a function
    of disc scale-length adapted from \citet{Ledo2010} and including
    the NSD of NGC~4698 \citep{Corsini2012}.  The large red square
    marks the NSD of model A2. Open circles indicate high
    surface-brightness spirals, triangles low surface-brightness
    spirals, stars show S0s, and filled circles discy ellipticals.
    Small squares represent observed NSDs.}}
\label{fig:discs}
\end{figure}

We compared the structural properties of the NSD of model A2 to those
of the known NSDs \citep{Ledo2010, Corsini2012}.
We derive the Johnson $V$-band central surface-brightness of the model
NSD from the mean $I$-band value using the {\sc IRAF} task {\sc
  SYNPHOT}.  Because this correction depends on the galaxy spectral
energy distribution, it is calculated using the spectral template for
a Sc spiral galaxy by \citet{Kinney1996} to match the late
morphological type of NGC~4244. The resulting shift is $V-I = 1.46$.
The mean observed central surface-brightness is corrected for
mean inclination as
\begin{equation}
\mu_{0,V}^0 \,=\, \mu_{0,V} -\,2.5 \log(\cos i )\,=\,20.24
\;{\rm mag\;arcsec}^{-2}.
\end{equation}
The scale-length and face-on central surface-brightness are close to
those of the larger observed NSDs and fit on the $\mu_{0,V}^0 - h$
relation for galaxy discs, as we show in Fig.~\ref{fig:discs}.

\begin{figure}
\center
\psfig{file=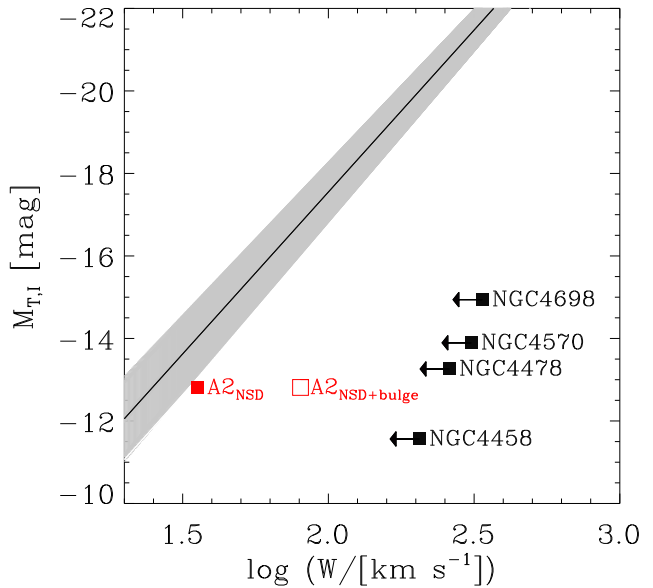,width=8.5cm,bbllx=115, bblly=250, bburx=510, bbury=610}
\caption{\small{The total magnitude as a function of the rotation
    width for the NSDs observed in real galaxies (black squares) and
    in the nucleus of model A2 (red squares). Only observed NSDs with
    both measured kinematics and photometric decomposition are shown.
    The two values for model A2 are derived by excluding (filled red
    square) and including (open red square) the bulge contribution.
    The solid line and the grey region correspond to the Tully-Fisher
    relation in the $I$ band by \citet{Masters2006} and its scatter,
    respectively.}}
\label{fig:TF}
\end{figure}

We also compare the total luminosity of the NSD with that predicted by
extrapolating the $I$-band Tully-Fisher relation
\citep[][Fig.~\ref{fig:TF}]{Masters2006} to discs of similar
rotational velocity ($\sim20$ km~s$^{-1}$). We assume the rotation
width of the NSD to be $W\,=\,2 \langle V_{\rm rms} \rangle$ where the
luminosity-weighted second-order kinematic moment $\langle V_{\rm rms}
\rangle \,=\, \sqrt{\langle V^2 \rangle + \langle \sigma^2 \rangle}$
is measured from the kinematic maps of the simulated galaxy excluding
the light contribution of the bulge (Table~\ref{fig:kinematics}). We
find that the rotation width and luminosity of the NSD of model A2 are
consistent with the Tully-Fisher relation.
Similarly we measure the rotational width for the NSDs for which both
the stellar kinematics and photometric decomposition are available so
far (NGC~4458 and NGC 4478: kinematics from \citealt{Halliday2001} and
photometry from \citealt{Morelli2004}; NGC~4570: kinematics from
\citealt{Krajnovic2004} and photometry from \citealt{Morelli2010};
NGC~4698: kinematics from \citealt{Bertola1999b} and photometry from
\citealt{Corsini2012}). Their luminosities are offset below the
Tully-Fisher relation in agreement with the earlier findings of
\citet{Morelli2004}.
The NSDs in real and simulated galaxies show a different behaviour in
the Tully-Fisher relation in spite of having similar luminosities and
similar maximum light contributions (Fig.~\ref{fig:TF}). We explain
the discrepancy as due to the fact that for real galaxies observations
miss a proper decomposition of the line-of-sight velocity
distribution. The rotation widths of NSDs in real galaxies are
over-estimated due to the contribution of the host spheroid to the
kinematics (in particular to the velocity dispersion) measured in the
nuclear regions.
This is not the case of the NSD of model A2 where we have the actual
kinematics of the nuclear disc, making the comparison with the
Tully-Fisher prediction straightforward. Its rotational width
increases and the NSD drops below the Tully-Fisher relation if the
measured stellar kinematics includes the bulge contribution. However,
then the rotational width ($\simeq80$ km~s$^{-1}$) is not as large as
in the observed NSDs ($\simeq230$ km~s$^{-1}$) since its bulge has a
lower velocity dispersion compared to the larger spheroidal component
of the early-type galaxies hosting the NSDs shown in
Fig.~\ref{fig:TF}.

\section{Conclusions}
\label{sec:conclusions}

We have analysed two $N-$body simulations from H$11$ exploring the
dissipationless merging of multiple star clusters into the centre of a
galaxy.  The simulations were originally aimed at investigating the
photometric, kinematic, and dynamic properties of the nucleus of
NGC~4244 which hosts a massive stellar cluster in rapid rotation
\citep{Seth2006, Seth2008}. In this paper we have investigated the
images and kinematic maps built from the simulation as if they were
real, assuming the galaxy to be at the distance of the Virgo cluster.
We have tested the importance of purely stellar dynamical mergers for
the formation and growth of NSDs by looking for their presence in the
nucleus of the simulated galaxies.
Our main conclusions can be summarised as follows.

\begin{itemize}

\item A flattened merger remnant ($q\simeq0.3-0.4$) with a radius of
  about 100 pc is observed in the nucleus of the simulated galaxy when
  a few tens of star clusters with sizes and masses comparable to
  those of globular clusters observed in the Milky Way are accreted
  onto a pre-existing stellar component at the centre. This flattened
  structure forms regardless of whether the pre-existing component is
  a massive spherical cluster (as is the NCS of model A1) or a rapidly
  rotating disc (as is the NCD of model A2) and regardless of the
  amount of accreted mass ($2-10\times$) with respect to the mass of
  the pre-existing component ($\sim 10^6$ M$_\odot$).

\item The merger remnant passes all observational constraints to be a
  NSD when the star clusters are accreted onto a pre-existing NCD. The
  structural parameters of the NSD were obtained by applying the same
  photometric decomposition adopted for real galaxies based on the
  analysis of isophotal shapes. The photometric and kinematic
  properties of the NSD in the simulated galaxy are remarkably similar
  to those of NSDs observed in the nuclei of real galaxies. In
  particular, the scale-length (38.8 pc) and face-on central
  surface-brightness (20.24 $V$-mag arcsec$^{-2}$) fit on the
  $\mu_{0,V}^0\,-\,h$ relation for galaxy discs. The total luminosity
  ($5.3 \times 10^6$ L$_\odot$) is consistent with that predicted by
  extrapolating the $I$-band Tully-Fisher relation to discs of similar
  rotational width ($\sim40$ km~s$^{-1}$). The mass of the NSD ($\sim
  1 \times 10^7$ M$_\odot$) is close to that of the few NSDs for which
  it has been measured.

\item The independent analysis performed by fitting an exponential
  disc to the surface-brightness distribution after excluding the
  light from the bulge finds best-fitting disc parameters in agreement
  with those derived with the Scorza-Bender method, showing that the
  latter is robust.  The elongated structures found in the unsharp
  images are not sufficient proof of the presence of a NSD: the same
  feature can also be caused by isophotes of increasing inwards
  ellipticity at small radii. Therefore, to unveil a NSD, performing a
  detailed measurement of the surface-brightness distribution in the
  nuclear regions is necessary and the Scorza-Bender decomposition
  method is a robust way of doing this.

\item The purely stellar dynamical merger of star clusters onto the
  centre of a galaxy is a viable mechanism for growing a NSD.  This
  shows that most of its mass (up to $70-90$ per cent) can be
  assembled from already formed stars through the migration and
  accretion of star clusters onto the galactic centre.

\end{itemize}

\section*{Acknowledgments}

E.P. acknowledges the Jeremiah Horrocks Institute of the University of
Central Lancashire in Preston, UK, for the hospitality while this
paper was in progress. E.P. is partially supported by Fondazione
Angelo Della Riccia. L.M. acknowledges Padua University grant
CPS0204. This work was supported by Padua University through the
grants 60A02-1283/10, 5052/11, and 4807/12.  V.P.D. is supported in
part by STFC Consolidated grant \# ST/J001341/1.  Simulations in this
paper were performed on the COSMOS Consortium supercomputer within the
DIRAC Facility jointly funded by STFC, the Large Facilities Capital
Fund of BIS and the High Performance Computer Facility at the
University of Central Lancashire.

\label{lastpage}

\end{document}